\documentclass[%
 reprint,
 amsmath,amssymb,
 prb,
]{revtex4-1}
\usepackage{graphicx}
\usepackage{dcolumn}
\usepackage{bm}

\begin{document}

\preprint{APS/123-QED}

\preprint{APS/123-QED}
\title{Effects of correlations on phase diagrams of the two-dimensional Su-Schrieffer-Heeger model with the longer-range hoppings} 
\author{Tao Du}
\email{Corresponding author: dutao@ymu.edu.cn}
\author{Yuexun Li}
\author{Helin Lu}
\author{Hui Zhang}
\affiliation{Department of Physics, Yunnan Minzu University, Kunming 650504, P. R. China}

\begin{abstract}
For the  Su-Schrieffer-Heeger (SSH) model on the two-dimensional square lattice, two third nearest neighbor hoppings which preserve chiral symmetry are introduced. Like the case of one dimension, the longer-range hopping can drive topological transitions and leads to the larger topological invariant (vectored Berry phase for the two dimensional SSH model). We obtain the phase boundaries and topological phase diagrams from the winding pattern of $\rho(k_l)$ of the Bloch Hamiltonian. Effects of correlations on the extended two dimensional SSH model are also investigated. The correlation shifts the phase boundaries and leads to topological transitions. Several special points in non-interacting phase diagrams are chosen to illustrate the different phase transitions and a correlations-driven topologically non-trivial phase from the trivial one is found. In this work, the slave-rotor mean field method is applied to the interacting model and we focus on the case of physical electrons, i.e. the Mott transition of charge degree of freedom is avoided in our investigations. 
\begin{description}
\item[Key words]2D SSH model, longer-range hoppings, correlations-driven transition,slave-rotor mean field method.
\item[PACS numbers]73.20.At, 71.27.+a, 71.30.+h, 71.10.Fd

\end{description}
\end{abstract}

\maketitle


\section{\label{sec1:level1}Introduction}
A simple one-dimensional model of topological insulators is the Su-Schrieffer-Heeger (SSH) model which was originally supposed to describe the soliton formation in polyacetylene \cite{1979Su}. It possesses the bipartite lattice structure and its topological nature is characterized by the Zak phase \cite{1989Zak} --- the Berry phase\cite{1984Berry} in the Brillouin zone. The Zak phase is closely connected with the winding number of vector $\bm{d}(k)$ in the Bloch Hamiltonian ${\mathcal H}_k=\bm{d}(k)\cdot \bm{\sigma}$.\cite{2016Asboth} When inter-cell hopping amplitude is larger than the intra-cell amplitude, the Zak phase is non-trivial, actually equal to $\pi$, and there is a zero-energy edge mode for open boundary conditions\cite{2011Delplace,2012Shen,2016Asboth}. 

Several extensions of the SSH model have been  provided by some authors. For example, one of these extensions is the introduction of the longer-range hoppings \cite{2014Li,2018Maffei,2019Perez-Gonzalez,2020Ahmadi,2020Hsu}. These longer-range hoppings can be classified into  two distinct classes. The ``odd hoppings" which connect sites of different sublattices (e.g. the third-nearest neighbor hopping) preserve the chiral symmetry of original SSH model and a quantized Zak phase can be applied to characterize the topologically non-trivial phases. The ``even hoppings" which connect sites of same sublattices (e.g. the next-nearest neighbor hopping) break the chiral symmetry and the model in this case is topologically trivial. At present, the main focus of interest is the emergence of the new phase with the larger Zak phase caused by the (odd) longer-range hoppings. Chen and Chiou \cite{2020Chen} have already given a proof that arbitrary odd longer-range hoppings can lead to \textit{arbitrary} winding numbers of Bloch vector $\bm{d}(k)$ (hence arbitrarily large Zak phase). The introduction of the spin-orbit coupling (SOC) which preserve the chiral symmetry was also considered for the spinful SSH model \cite{2020Ahmadi,2014Yan,2016Bahari,2017Yao,2020Bahari}. Yan and Wan \cite{2014Yan}, and Bahari and Hosseini \cite{2016Bahari} showed that the nearest neighbor SOC lifts the spin degeneracy of Bloch bands and leads to new behaviors of the closing and reopening of gaps. Then, although the largest Zak phase of non-trivial phases remains unchanged, the more topological phases and topological transitions will occur. Ahmadi and Abouie \cite{2020Ahmadi} investigated the interplay of nearest neighbor SOCs and the longer-range hoppings. The complex phase diagram they obtained shows that, besides topologically non-trivial phases with the largest Zak phase caused by the longer-range hoppings, there are several phases with non-trival Zak phases caused by the SOC. Although the mechanism of the emergence of these phases was not emphasized in their work, it is easy to realize that it should be the same as that pointed out in Ref.~\onlinecite{2014Yan}. Another extension is that the unit cell of SSH model can be enlarged to form the superlarttice \cite{2015Guo,2018Maffei,2019Xie,2020Bahari}. Maffei \textit{et al.} \cite{2018Maffei}, Guo and Chen \cite{2015Guo}, and Xie \textit{et al.} \cite{2019Xie} have investigated the so-called SSH$_4$ model which possesses four sites in a unit cell. These investigations indicate the existence of topologically non-trivial phases with the larger Zak phase.   Furthermore, there is a relation between the \textit{winding number} (a special topological invariant which can, in principle, be connected to Zak phase) and the number of sublattices per unit cell  \cite{2020Bahari}. It astonishingly leads to the existence of a maximum winding number.

A two-dimensional extension of SSH model in space has been done on the square \cite{2017aLiu,2018Liu,2019Obana,2019Yuce} or honeycomb \cite{2017bLiu} lattice . These two-dimensional SSH models have zero Berry curvature everywhere in Brillouin zone, except at band gap closing points, due to inversion and time-reversal symmetries, and the Zak phase or Chern number is zero. However, there is also a topologically non-trivial phase characterized by a so-called two-dimensional (vectored) Zak phase \cite{2017aLiu,2012Fang}. In our work here, one of our interests is the extension of the two-dimensional SSH model on the square lattice by introducing the longer-range hoppings. We expect that the topologically non-trivial phases with the larger topological invariant (vectored Zak phase here) to emerge, similarly to the case of one dimension. Interestingly, there is a similarity between the two-dimensional SSH model and the Chern insulator model. That is, for Chern insulators the longer-range hopping can lead to the larger Berry phase or Chern number \cite{2012Sticlet,2013Sticlet}, while for two-dimensional SSH model it also leads to the larger vectored Zak phase or winding number. By inspecting the vectored Zak phase, we will obtain the boundaries between distinct phases and topological phase diagrams.   

Over the past decade, the investigation of effects of correlations, especially strong correlations, on topological insulators has been a main topic in condensed matter physics \cite{2013Hohenadler,2014Imada,2018Rachel}. For one-dimensional spinful SSH model, when the non-interacting SSH model possesses the topologically non-trivial phase, the arbitrary on-site Hubbard interaction can adiabatically lead to a topological Mott insulator \cite{2012Manmana}. The result is based on the analysis of a topological invariant constructed from Green's functions, and of the time evolution of the local spin density (i.e. a local spin emerges at boundaries). Later, the topologically non-trivial Mott phase was also confirmed by the degeneracy of the ground entanglement spectrum \cite{2016Ye,2014Yoshida}. Moreover, Yoshida \textit{et al.} \cite{2014Yoshida} showed a correlations-driven edge-Mott state with gapful charge and gapless spin excitations, which corresponds to the degeneracy of the entanglement spectrum of bulk. For the spinless SSH model, the correlation represented by the nearest neighbor Hubbard interaction was supposed to induce the charge density wave state (CDW) in the whole domain of the nearest neighbor hoppings \cite{2018Yahyavi}. However, the effects of correlations on the two-dimensional SSH model have not yet been investigated. In this work, we introduce the on-site Hubbard interaction into our extended two-dimensional SSH model and deal with the interacting model by the slave-rotor mean field method \cite{2002Florens,2003Florens,2004Florens}. The main goal here is to investigate influences of the correlation on phase boundaries and the topological phase diagram, and to find the correlation-driven topological transitions. We shall focus on the \textit{physical electron}, and then, in the slave-rotor mean field treatment, restrict the strength of Hubbard interactions to maintain all of the phases below the Mott transition of charge degree of freedom. Unlike the treatments of one dimensional interacting SSH models given by other authers \cite{2012Manmana,2014Yoshida}, where the topological invariant is extracted from the Green's function, in our mean field treatment the effect of correlations is reflected in the \textit{renormalization} of model parameters. From these \textit{renormalized} model parameters, we can obtain the vectored Zak phase of interacting SSH model in the same way as in the non-interacting case. So, it is easier to obtain topological phases and topological transitions. 

This paper is organized as follows. In sec. \uppercase\expandafter{\romannumeral2}, we give the extended two-dimensional SSH model with longer-range hoppings and obtain the non-interacting topological phase diagrams.  In sec. \uppercase\expandafter{\romannumeral3}, the on-site Hubbard interaction is introduced into the model and the influences of correlations on the phase boundaries and topological phase diagram are investigated. The correlations-driven topological transitions are also studied in this section. Finally, we conclude in sec. \uppercase\expandafter{\romannumeral4}.

\section{\label{sec2}The two-dimensional SSH model with the longer-range hoppings}

\subsection{\label{sec2-1}The model}

\begin{figure}[ht]
\includegraphics[width=6cm,height=6cm]{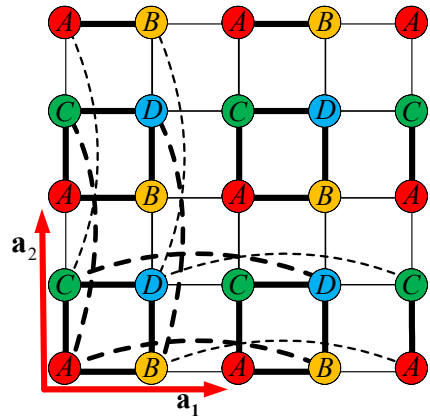}
\caption{\label{fig1}(Color online)The lattice structure of L-SSH model. Red solid arrows represent the primitive vectors ${\bm a}_{1}=(a, 0)$ and ${\bm a}_{2}=(0, a)$. The lattice is constituted by sublattices A, B, C and D. Thick and thin solid lines denote intra- and inter-cell electron hoppings with strength $w$ and $v$ respectively. The thick and thin dashed lines denote the third nearest neighbor electron hoppings with strength $t_{12}$ and $t_{21}$ respectively.}
\end{figure}

The Hamiltonian of SSH model with longer-range hoppings (L-SSH) on the two-dimensional square lattice is
\begin{eqnarray}
H_{0}&=&\sum_{\langle ii^{\prime}\rangle j}\sum_{\sigma}t_{ix}\hat{c}^{\dagger}_{ij\sigma}\hat{c}_{i^{\prime}j\sigma}+\sum_{i\langle jj^{\prime}\rangle}\sum_{\sigma}t_{jy}\hat{c}^{\dagger}_{ij\sigma}\hat{c}_{ij^{\prime}\sigma}\nonumber\\
&&+\sum_{\{ii^{\prime}\}j}\sum_{\sigma}t_{ix}^{\prime}\hat{c}^{\dagger}_{ij\sigma}\hat{c}_{i^{\prime}j\sigma}+\sum_{i\{jj^{\prime}\}}\sum_{\sigma}t_{jy}^{\prime}\hat{c}^{\dagger}_{ij\sigma}\hat{c}_{ij^{\prime}\sigma}.
\label{eq1}
\end{eqnarray}
Here $\hat{c}_{ij\sigma}^{\dagger}$ ($\hat{c}_{ij\sigma}$) is the creation (annihilation) operator of an electron with spin $\sigma=\uparrow(\downarrow)$ at site ($i$,$j$). The first two terms are the nearest neighbor electrons hopping terms with the hopping strength $t_{ix}=t_{jy}=w$ when $i$ ($j$) is odd and $t_{ix}=t_{jy}=v$ if $i$ ($j$) is even. The last two terms represent the electron hoppings between the third nearest neighbor sites along the horizontal and vertical direction respectively, which can preserve the chiral symmetry. The hopping strength $t^{\prime}_{ix}=t^{\prime}_{jy}=t_{12}$ if $i$ ($j$) is odd while $t^{\prime}_{ix}=t^{\prime}_{jy}=t_{21}$ if even. The primitive cell should contain four lattice sites due to the type of the electron hoppings, as shown in Fig.~\ref{fig1}. 

The Hamiltonian of the extended SSH model can be transformed to momentum space as
\begin{eqnarray}
H_{0}=\sum_{\bm k}\begin{pmatrix}\Psi_{\bm k\uparrow}^{\dagger}&\Psi_{\bm k\downarrow}^{\dagger}\end{pmatrix}\begin{pmatrix}{\cal H}_{0\bm k\uparrow}&0\\0&{\cal H}_{0\bm k\downarrow}\end{pmatrix}\begin{pmatrix}\Psi_{\bm k\uparrow}\\\Psi_{\bm k\downarrow}\end{pmatrix}.
\label{eq2}
\end{eqnarray}
Here $\Psi_{\bm k\sigma}=\begin{pmatrix}\hat{c}_{\bm k\sigma}^{A}&\hat{c}_{\bm k\sigma}^{B}&\hat{c}_{\bm k\sigma}^{C}&\hat{c}_{\bm k\sigma}^{D}\end{pmatrix}^T $ is the matrix of electron operators in the momentum-spin space and the Bloch Hamiltonian is
\begin{eqnarray}
{\mathcal H}_{0\bm k\sigma}=
\begin{pmatrix}
0&\rho(k_{x})&\rho(k_{y})&0\\
\rho^{*}(k_{x})&0&0&\rho(k_{y})\\
\rho^{*}(k_{y})&0&0&\rho(k_{x})\\
0&\rho^{*}(k_{y})&\rho^{*}(k_{x})&0
\end{pmatrix},
\label{eq3}
\end{eqnarray}
where A, B, C, D represent the sublattices of the lattice as shown in Fig.~\ref{fig1} and $\rho(k)=w+ve^{-\mathrm{i}k}+t_{12}e^{\mathrm{i}k}+t_{21}e^{-\mathrm{i}2k}=|\rho(k)|e^{\mathrm{i}\phi(k)}$. It is easy to obtain the four eigenvectors of the Bloch Hamiltonian for each spin sectors as $\bm u_{0\sigma}=(1/2)\begin{pmatrix}1&s_{1}e^{-\mathrm{i}\phi(k_{x})}&s_{2}e^{-\mathrm{i}\phi(k_{y})}&s_{1}s_{2}e^{-\mathrm{i}\phi(k_{x})-\mathrm{i}\phi(k_{x})}\end{pmatrix}^{T}$ with the eigenvalues $\lambda=s_{1}|\rho(k_{x})|+s_{2}|\rho(k_{y})|$, here $s_{1}=\pm1$ and $s_{2}=\pm1$. We label the four bands of electrons as $(s_1,s_2)$.

\subsection{\label{sec2-2}The topological phase diagram}

The Chern number is equal to zero from the parity of Berry curvature for each band in the Brillouin zone (BZ), i.e. $F_{ij}(-k_{x},-k_{y})=-F_{ij}(k_{x},k_{y})$, due to the time-reversal (TR) symmetry  possessed by the system \cite{2013Bernevig}. The inversion symmetry can further cause the Berry curvature to vanish in the BZ except those locations of band degeneracy \cite{2017aLiu}, because this symmetry causes the Berry curvature to be a even function, i.e. $F_{ij}(-k_{x},-k_{y})=F_{ij}(k_{x},k_{y})$. Then the vectored Berry phase (the Zak phase is called Berry phase here and below) defined as follows has to be used to characterize topological properties of our extended 2D SSH \cite{2017aLiu,2012Fang}. That is
\begin{eqnarray}
\bm{\mathcal{C}}=\frac{1}{2\pi}\iint_{BZ}dk_{x}dk_{y}\mathrm{Tr}(\bm{A}).
\label{eq4}
\end{eqnarray} 
Here $\bm{A}=\mathrm{i}\bm u^{*}_{i}(\bm k)\bm{\nabla}_{k}\bm u_{j}(\bm k)$, $\bm u_{i}(\bm k)$ is the eigenvector of the $i$-th band of the Bloch Hamiltonian, and the ``$\mathrm{Tr}$" denotes the trace over bands. In particular, each component of the vectored Berry phase, i.e.
\begin{eqnarray}
\mathcal{C}_{l}=\sum_{i}^{occ.}\mathrm{i}\int_{-\pi}^{\pi}\int_{-\pi}^{\pi}dk_xdk_y\bm u^{*}_{i}(\bm k)\frac{\partial}{\partial k_{l}}\bm u_{i}(\bm k),
\label{eq5}
\end{eqnarray} 
can be calculated from the eigenvectrors $\bm{u}_{0\sigma}$ of Bloch Hamiltonian (\ref{eq3}). It is then straightforward to obtain these two components of the vectored Berry phase as
\begin{eqnarray}
\mathcal{C}_{l}=N_{occ}\cdot\frac{1}{2}\int_{-\pi}^{\pi}d\phi(k_{l}).
\label{eq6}
\end{eqnarray} 
In the two formulae above, the component of vectored Berry phase is given by the sum over all of the occupied bands, the number $N_{occ}$ of occupied bands includes the spin degree of freedom of electrons, and $\phi(k_{l})$ is the argument of $\rho({k_{l}})$, i.e.
\begin{eqnarray}
\phi(k_{l})=\arctan\frac{(-v+t_{12})\sin({k_{l})-t_{21}\sin(2k_{l})}}{w+(v+t_{12})\cos(k_{l})+t_{21}\cos(2k_{l})}.
\label{eq7}
\end{eqnarray} 
Then, the vectored Berry phase can be obtained from the winding number of the $\rho(k_{l})$ when it winds around the origin of the complex plane as momentum $k_{l}$ traverses the Brillouin zone along the $l$-direction. Firstly, we obtain critical conditions under which the winding number of $\rho(k_l)$ changes. It is the situation that $\rho(k_l)$ passes through the origin of the complex plane. From $\rho(k_{l})=0$, the conditions are obtained as
\begin{eqnarray}
w-v-t_{12}+t_{21}=0,
\label{eq8}\\
w+v+t_{12}+t_{21}=0,
\label{eq9}
\end{eqnarray} 
and
\begin{eqnarray}
t_{21}(w-t_{21})=t_{12}(v-t_{12})
\label{eq10}
\end{eqnarray} 
in the case of $k_{l}=\pm\pi$, $0$ and $\arccos[(t_{12}-v)/2t_{21}]$ respectively. 
\begin{figure}[ht]
\includegraphics[width=4.25cm,height=3cm]{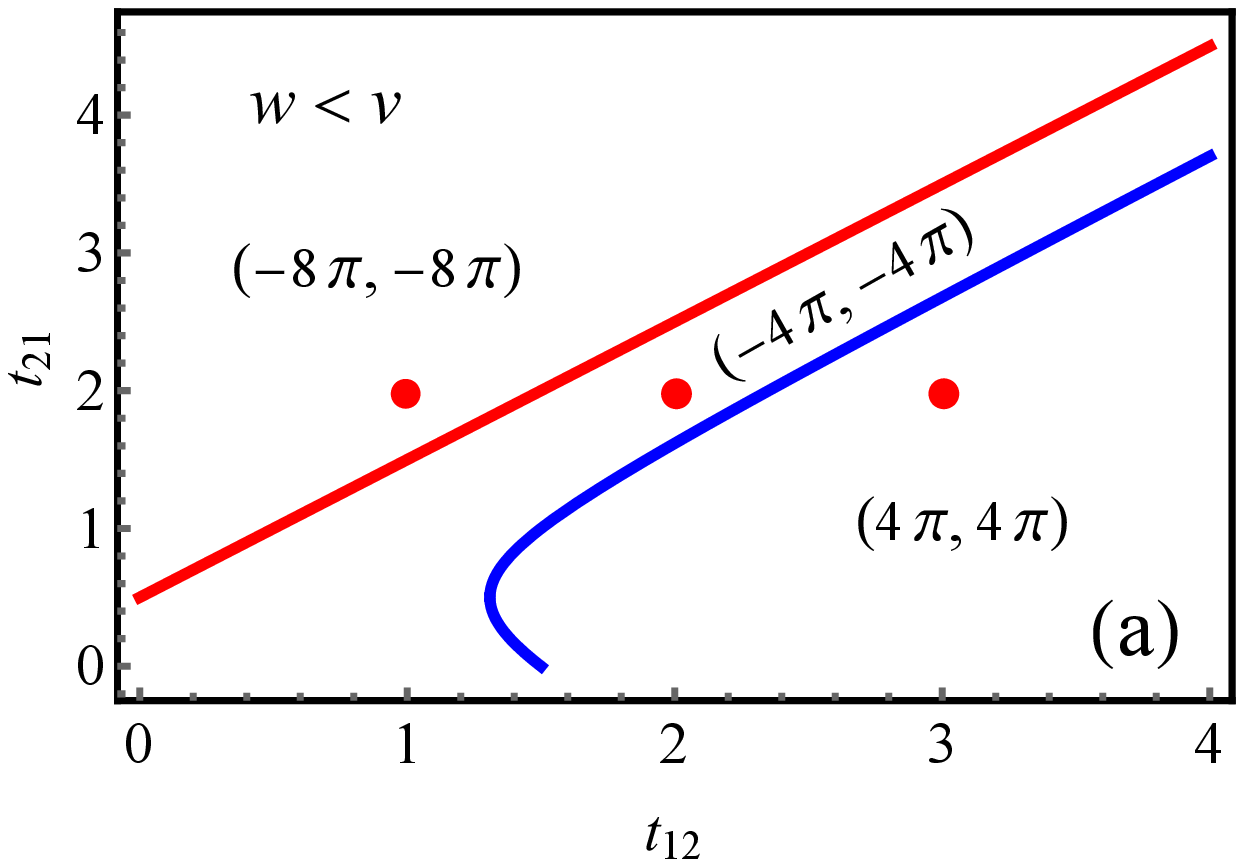}
\includegraphics[width=4.25cm,height=3.4cm]{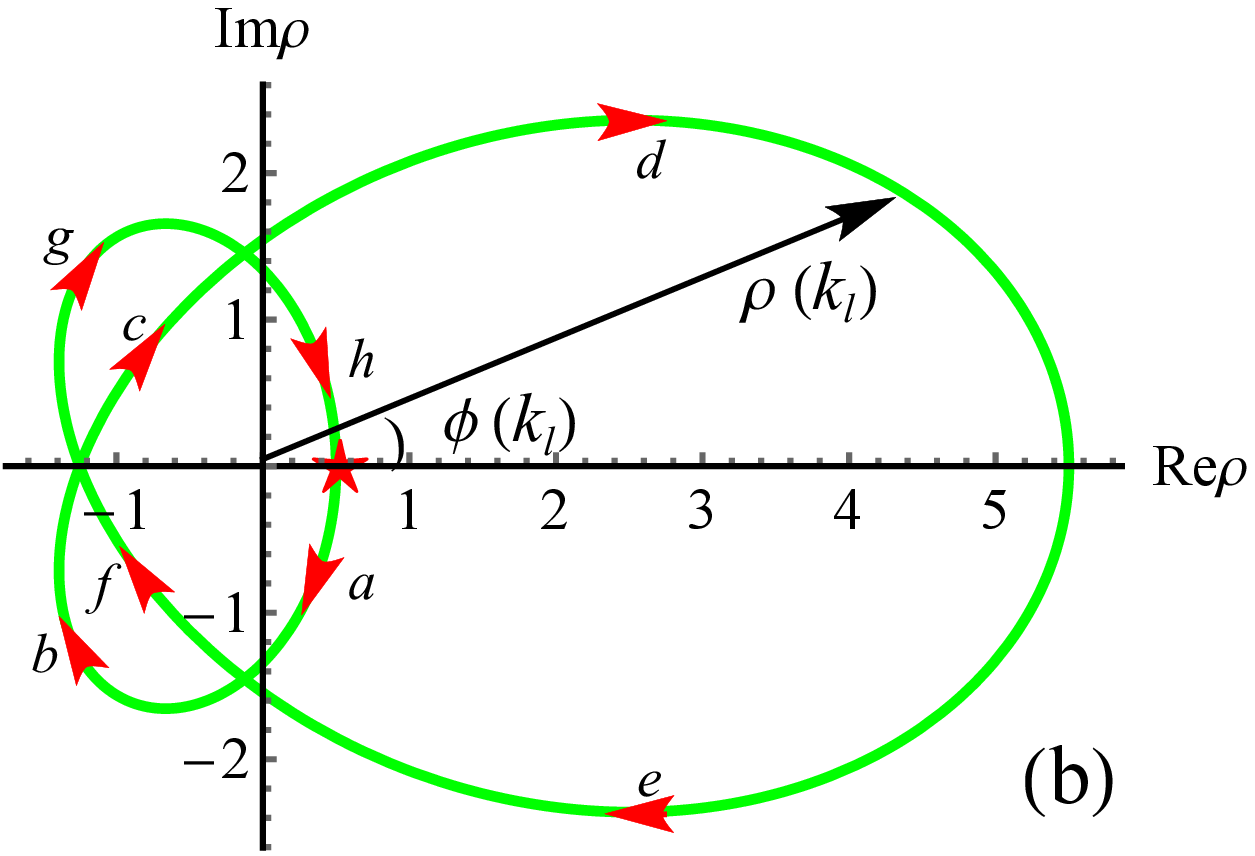}
\includegraphics[width=4.25cm,height=3.1cm]{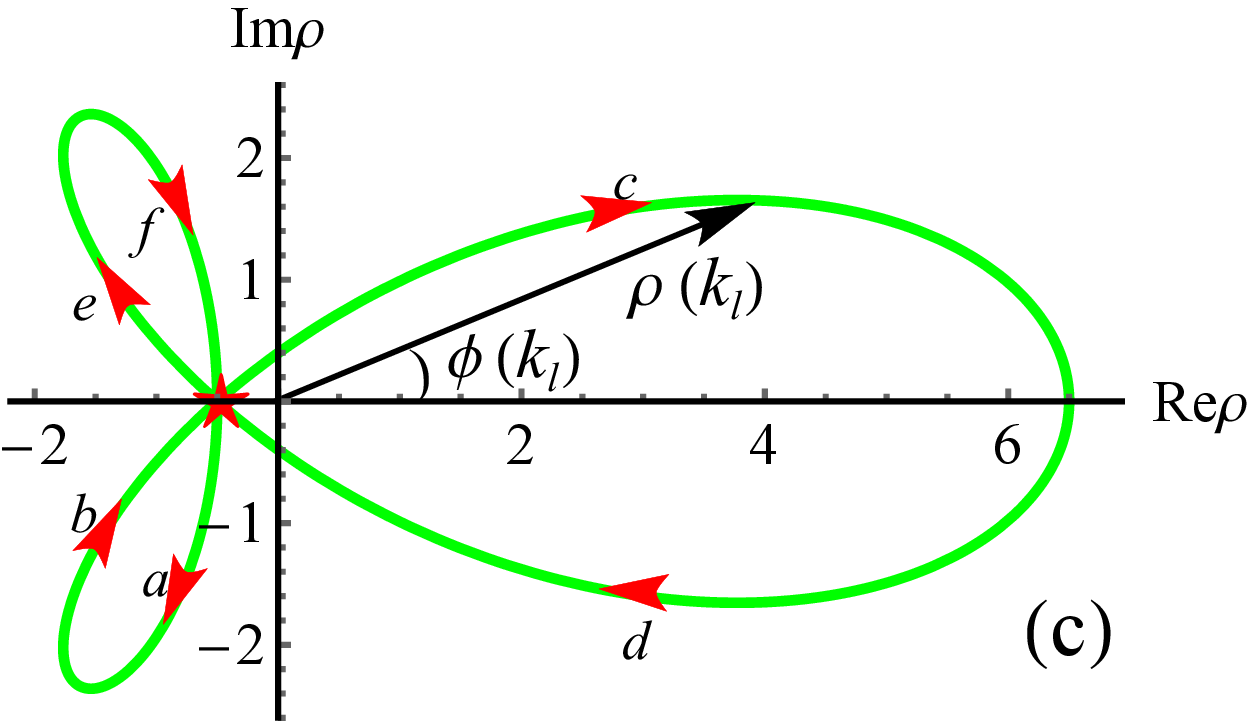}
\includegraphics[width=4.25cm,height=3.1cm]{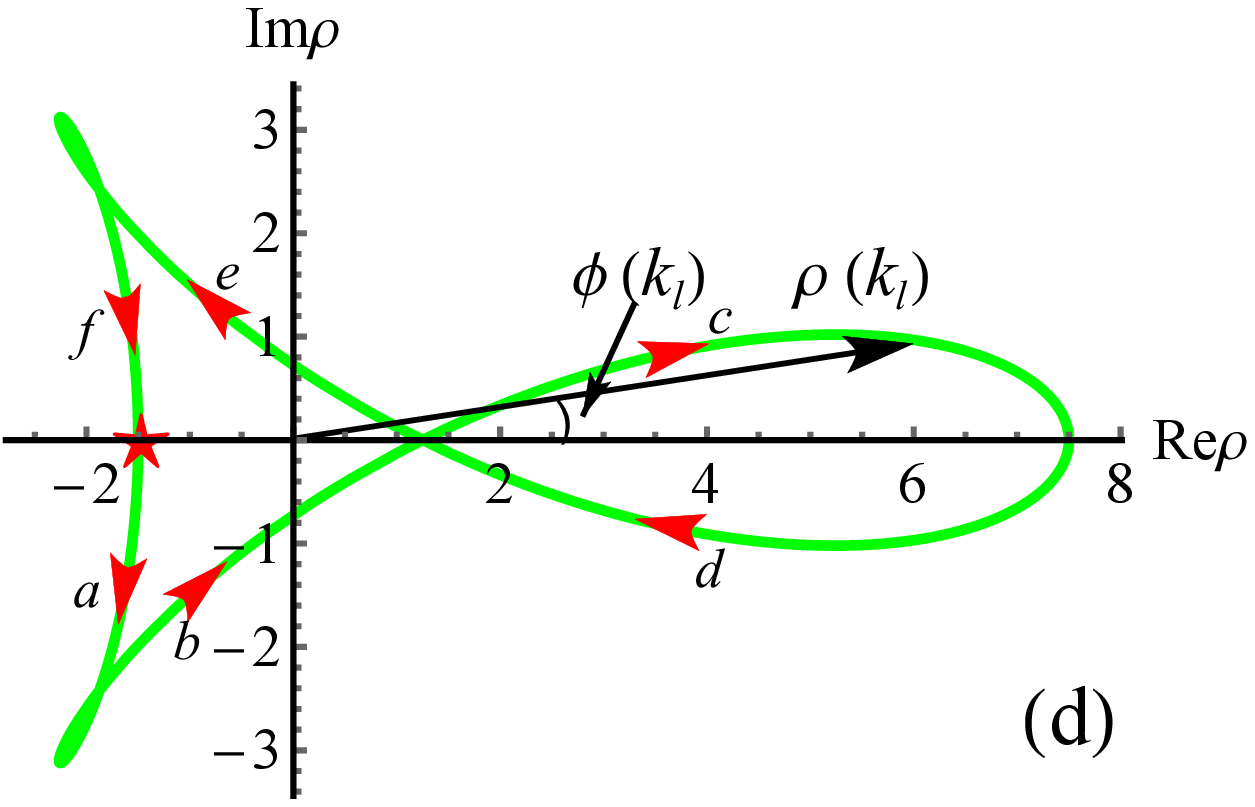}
\caption{\label{fig2}(Color online)(a)Topological phase diagram of the model at $w=1$ and $v=1.5$ in the case of half-filling. The red and blue lines are the boundaries at which the winding number of $\rho(k_{l})$ changes. The three regions correspond to vectored Berry phase--$(-8\pi, -8\pi)$, $(-4\pi, -4\pi)$ and $(4\pi, 4\pi)$ respectively. The trajectories that $\rho(k_{l})$ winds around origin in the case of $(t_{12},t_{21})=$ $(1,2)$, $(2,2)$ and $(3,2)$ (red points in (a)) are shown in (b), (c) and (d) respectively. Here, red arrowheads and letters a, b,... represent the direction of the flow of $\rho(k_{l})$. The red star represents the starting (terminal) point of the trajectory at $k_{l}=-\pi(\pi)$.}
\end{figure}
The boundaries at which the winding number of $\rho(k_{l})$, i.e. the vectored Berry phase of the model changes are shown in Fig.~\ref{fig2}(a) in the case of $w<v$ and $t_{12}(t_{21})>0$. The red line is obtained from Eq.~(\ref{eq8}) and the blue curve from Eq.~(\ref{eq10}). The vectored Berry phase of the bands for each region of $t_{12}$-$t_{21}$ plane can be obtained by observing how $\rho(k_{l})$  winds around the origin of complex plane. There are three different winding pattern as shown in Fig.~\ref{fig2}(b)-(d). Correspondingly, in the case of half-filling, the vectored Berry phase $\bm{\mathcal{C}}$ are obtained as $(-8\pi, -8\pi)$, $(-4\pi, -4\pi)$ and $(4\pi, 4\pi)$ when $\phi(\pi)-\phi(-\pi)$ are equal to $-4\pi$, $-2\pi$ and $2\pi$ respectively. Then, we get the topological phase diagram of the extended 2D SSH model at $w<v$ (Fig.~\ref{fig2}(a)). In the phase diagram, the point $(t_{12},t_{21})=(0,0)$ corresponds to the conventional 2D SSH model which possesses topologically non-trivial ground states with vectored Berry phase $(-4\pi, -4\pi)$. It can be seen here that the longer-range hopping changes and even enlarges the topological invariant. The model studied here is also an illustration of the mechanism that the longer-range hopping can enlarge the topological invariant.   

\begin{figure}[ht]
\includegraphics[width=6.5cm,height=4.5cm]{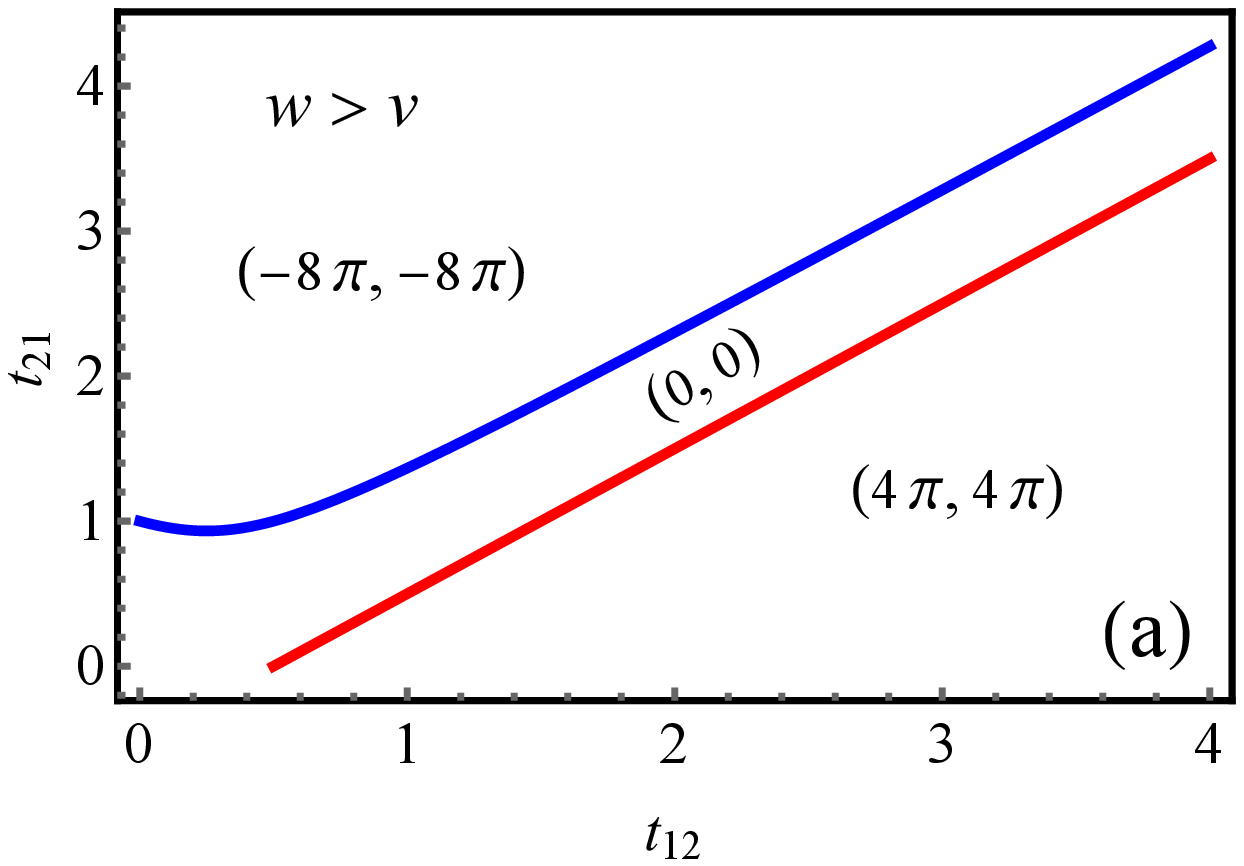}
\includegraphics[width=6.5cm,height=4.5cm]{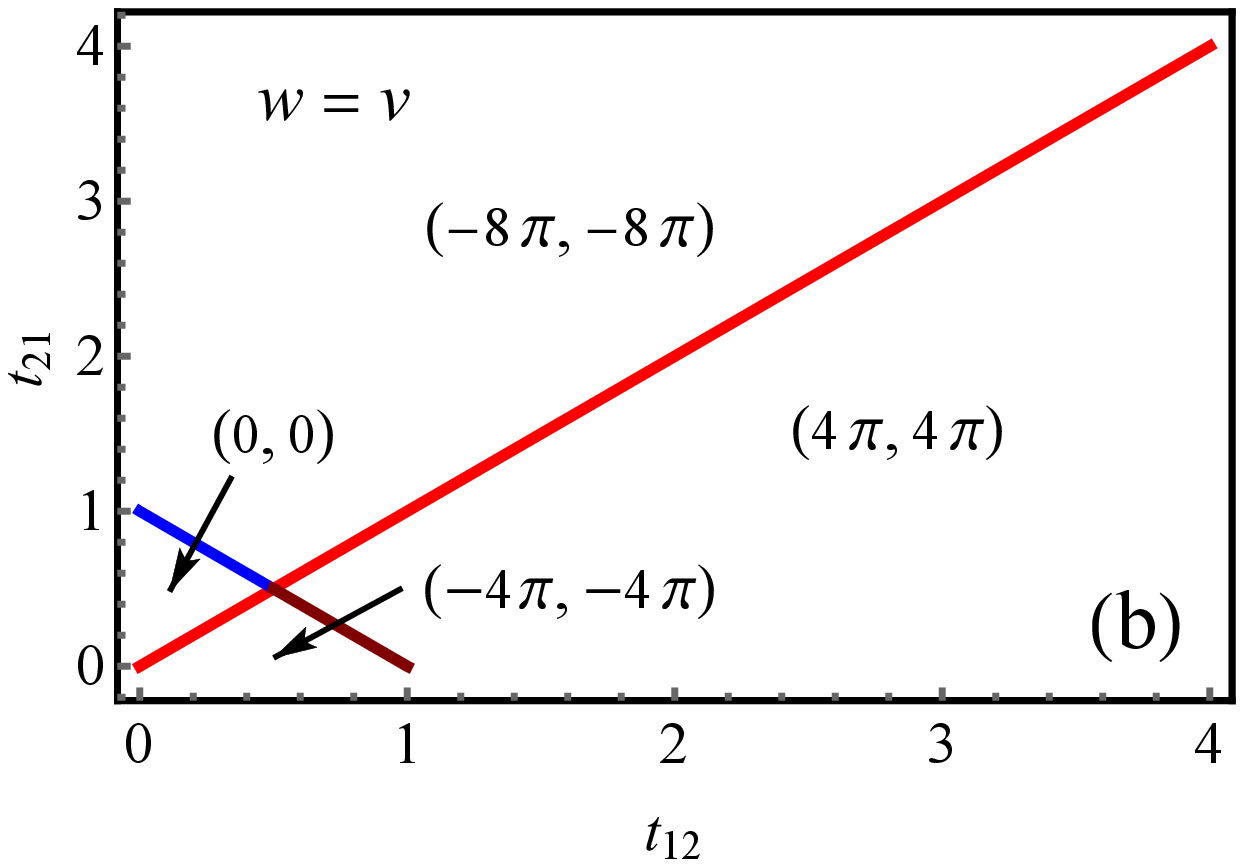}
\caption{\label{fig3}(Color online)Topological phase diagrams when (a) $w=1,v=0.5$ and (b) $w=1,v=1$ in the case of half-filling. The red and blue lines are obtained from Eq.~(\ref{eq8}) and (\ref{eq10}) respectively. The vectored Berry phase of each sector can also be obtained from investigations on how $\rho(k_{l})$  winds around the origin of complex plane.}
\end{figure}

When $w>v$, the corresponding conventional SSH model is topologically trivial, which possesses vectored Berry phase $(0, 0)$. The phase diagram of the extended SSH model in this case is shown in Fig.~\ref{fig3}(a). It shows that the topological invariant will not change if strengths of the two types of longer-range hopping are roughly equal. When one of them is larger, it can drive a topological transition and lead to topologically non-trivial phase with the larger topological invariant. When $w=v$, the conventional SSH model is just a tight-binding model with the nearest neighbor hopping in two dimension. It is the critical point at which the system changes from a topologically trivial phase to a non-trivial one. The two longer-range hopping  also drive the system into some topologically non-trivial phase. The richer phase diagram is shown in Fig.~\ref{fig3}(b).  

\section{\label{sec3}The L-SSH model with correlations}

\subsection{\label{sec3-1}The treatment from the slave-rotor mean field method}
The correlations can be represented by the on-site Hubbard interaction as
\begin{eqnarray}
H_{U}=\frac{U}{2}\sum_{i,j}\left(\sum_{\sigma}n_{ij\sigma}-1\right).
\label{eq11}
\end{eqnarray} 
Here $n_{ij\sigma}=\hat{c}^{\dagger}_{ij\sigma}\hat{c}_{ij\sigma}$ is the number operator of electrons with spin-$\sigma$ at the site $(i,j)$. 

An effective approach to the Hubbard model at half-filling is the slave-rotor method\cite{2002Florens,2003Florens,2004Florens}. In the method, the physical electron in the correlated system can be viewed as a composite of chargons and spinons. The Mott transition of strongly correlated systems is characterized by the transition of the chargon where the condensed chargon becomes the uncondensed one. Below the Mott boundary, the condensed chargon combines the spinon to form the conventional physical electron. The advantageous feature is that the spinon inherits the band structure of physical electrons, but with the renormalization of model parameters which stems from the correlation. Then, we can treat the band structure of correlated electrons through spinons, as the case of uncorrelated electrons. In the slave-rotor method the physical electron annihilation operator is decomposed as $\hat{c}_{ij\sigma}=e^{\mathrm{i}\theta_{ij}}\hat{f}_{ij\sigma}$. Here $e^{\mathrm{i}\theta_{ij}}$ is the U(1) rotor operator that describes the chargons and $\hat{f}_{ij\sigma}$ is the spinon operator. There is a constraint as $\sum_{\sigma}\hat{f}_{ij\sigma}^{\dagger}\hat{f}_{ij\sigma}+ \hat{L}_{ij}=1$, which is introduced to recover the Hilbert space of the physical electron, where the canonical angular momentum $\hat{L}_{ij}=\mathrm{i}\partial_{\theta_{ij}}$ associated with the angular $\theta_{ij}$. In the decomposition of physical electrons, the Hamiltonian of the correlated L-SSH model i.e. $H=H_{0}+H_{U}$, can be rewritten as
\begin{eqnarray}
H^{S}&&=\!\sum_{\langle ii^{\prime}\rangle j\sigma}\!t_{ix}e^{-\mathrm{i}\theta_{ii^{\prime},j}}\hat{f}^{\dagger}_{ij\sigma}\hat{f}_{i^{\prime}j\sigma}+\!\sum_{i\langle jj^{\prime}\rangle\sigma}\!t_{jy}e^{-\mathrm{i}\theta_{i,jj^{\prime}}}\hat{f}^{\dagger}_{ij\sigma}\hat{f}_{ij^{\prime}\sigma}\nonumber\\
&&+\!\sum_{\{ii^{\prime}\}j\sigma}\!t_{ix}^{\prime}e^{-\mathrm{i}\theta_{ii^{\prime},j}}\hat{f}^{\dagger}_{ij\sigma}\hat{f}_{i^{\prime}j\sigma}+\!\sum_{i\{jj^{\prime}\}\sigma}\!t_{jy}^{\prime}e^{-\mathrm{i}\theta_{i,jj^{\prime}}}\hat{f}^{\dagger}_{ij\sigma}\hat{f}_{ij^{\prime}\sigma}\nonumber\\
&&+\frac{U}{2}\sum_{ij}\hat{L}_{ij}^{2}-\mu\sum_{ij\sigma}\hat{f}_{ij\sigma}^{\dagger}\hat{f}_{ij\sigma}.
\label{eq12}
\end{eqnarray} 
Here the first four terms are simply the $H_{0}$ in the Slave-rotor decompostion ($H^{S}_{0}$), in which $\theta_{ii^{\prime},j}=\theta_{ij}-\theta_{i^{\prime}j}$ and $\theta_{i,jj^{\prime}}=\theta_{ij}-\theta_{ij^{\prime}}$. The partition function of the system is written as a path integral of $e^{-S_{E}}$ over fields $f$, $f^{*}$ and $\theta$, where
\begin{eqnarray}
S_{E}&=&\int_{0}^{\beta}d\tau\Big[\sum_{ij}\mathrm{i}L_{ij}\partial_{\tau}\theta_{ij}+\sum_{ij\sigma}f^{*}_{ij\sigma}\partial_{\tau}f_{ij\sigma}+H^{S}\nonumber\\
&&+\sum_{ij}h_{ij}\Big(\sum_{\sigma}f^{*}_{ij\sigma}f_{ij\sigma}+L_{ij}-1\Big)\Big]
\label{eq13}
\end{eqnarray}
is the action in imaginary time $\tau=\mathrm{i}t$. Here the last term is introduced into the action to impose the constraint condition. From the canonical equation of motion in imaginary time, i.e. $\mathrm{i}\partial_{\tau}\theta_{ij}=-\partial H/\partial L_{ij}$, we can obtain the relation of $L$ and $\theta$ as $L_{ij}=-(\mathrm{i}/U)\partial_{\tau}\theta_{ij}$. Then the action can be obtained as 
\begin{eqnarray}
S_{E}&=&\int_{0}^{\beta}\!d\tau\Big[\sum_{ij\sigma}f^{*}_{ij\sigma}(\partial_{\tau}\!-\!\mu\!+\!h_{ij})f_{ij\sigma}\!+\!\sum_{ij}(-h_{ij}\!+\!\frac{h_{ij}^{2}}{2U})\nonumber\\
&&+\frac{1}{2U}\sum_{ij}(\partial_{\tau}\theta_{ij}-\mathrm{i}h_{ij})^{2}+H_{0}^{S}\Big]
\label{eq14}
\end{eqnarray}
A constrained (complex) bosonic field $X_{ij}=e^{\mathrm{i}\theta_{ij}}$ with $|X_{ij}|^2=1$ and four mean field parameters, i.e. 
\begin{eqnarray}
&&Q_{X}=\big<\sum_{\sigma}f_{ij\sigma}f_{i^{\prime}j\sigma}\big>_{\langle ii^{\prime}\rangle}=\big<\sum_{\sigma}f_{ij\sigma}f_{ij^{\prime}\sigma}\big>_{\langle jj^{\prime}\rangle},
\label{eq15}\\
&&Q_{f}=\big<X_{ij}^{*}X_{i^{\prime}j}\big>_{\langle ii^{\prime}\rangle}=\big<X_{ij}^{*}X_{ij^{\prime}}\big>_{\langle jj^{\prime}\rangle},
\label{eq16}\\
&&Q_{X}^{\prime}=\big<\sum_{\sigma}f_{ij\sigma}f_{i^{\prime}j\sigma}\big>_{\{ii^{\prime}\}}=\big<\sum_{\sigma}f_{ij\sigma}f_{ij^{\prime}\sigma}\big>_{\{jj^{\prime}\}},
\label{eq17}\\
&&Q_{f}^{\prime}=\big<X_{ij}^{*}X_{i^{\prime}j}\big>_{\{ii^{\prime}\}}=\big<X_{ij}^{*}X_{ij^{\prime}}\big>_{\{jj^{\prime}\}},
\label{eq18}
\end{eqnarray}
are introduced to let the action $S_{E}$ to be expressed in the quadratic form of $X$-field and $f$-field as follows:
\begin{eqnarray}
S_{E}&=&\!\int_{0}^{\beta}\!d\tau\Big[\frac{1}{2U}\!\sum_{i}\mathrm{i}\partial_{\tau}X_{ij}^{*}(-\mathrm{i}\partial_{\tau})X_{ij}\!+\!\sum_{i}\!\rho_{ij}|X_{ij}|^{2}\!+\!H^{X}\nonumber\\
&&+\sum_{i\sigma}f_{i\sigma}^{*}\partial_{\tau}f_{i\sigma}+H^{f}+\cdots\Big].
\label{eq19}
\end{eqnarray}
Here, the symbol $``\cdots"$ denotes constant terms of mean field decomposition and we have set $h_{ij}\equiv h=-\mu=0$ for half-filling at the mean field level. $\rho_{ij}$ is the Lagrange multiplier for constraint $|X_{ij}|^2=1$ and $\rho_{ij}\equiv\rho$. In the expression of the action (\ref{eq19}), 
\begin{eqnarray}
H^{X}&=&Q_{X}\sum_{\langle ii^{\prime}\rangle j}t_{ix}X_{ij}^{*}X_{i^{\prime}j}+Q_{X}\sum_{i\langle jj^{\prime}\rangle}t_{jy}X_{ij}^{*}X_{ij^{\prime}}\nonumber\\
&&+Q_{X}^{\prime}\!\sum_{\{ii^{\prime}\}j}\!t_{ix}X_{ij}^{*}X_{i^{\prime}j}+Q_{X}^{\prime}\!\sum_{i\{ jj^{\prime}\}}\!t_{jy}X_{ij}^{*}X_{ij^{\prime}}
\label{eq20}
\end{eqnarray}
and 
\begin{eqnarray}
H^{f}\!&=&Q_{f}\!\sum_{\langle ii^{\prime}\rangle j\sigma}t_{ix}f_{ij\sigma}^{*}f_{i^{\prime}j\sigma}+Q_{f}\!\sum_{i\langle jj^{\prime}\rangle\sigma}t_{jy}f_{ij\sigma}^{*}f_{ij^{\prime}\sigma}\nonumber\\
&&+Q_{f}^{\prime}\!\!\sum_{\{ii^{\prime}\}j\sigma}\!t_{ix}f_{ij\sigma}^{*}f_{i^{\prime}j\sigma}+Q_{f}^{\prime}\!\!\sum_{i\{jj^{\prime}\}\sigma}\!t_{jy}f_{ij\sigma}^{*}f_{ij^{\prime}\sigma}.
\label{eq21}
\end{eqnarray}
Next, by the Fourier transforms
\begin{eqnarray}
&&X_{m}(\tau)=\frac{1}{\sqrt{\beta N_{\Lambda}}}{\sum_{\bm{k},n}}'e^{\mathrm{i}(\bm{k}\cdot\bm{R}_{m}-v_{n}\tau)}X_{\bm{k}}(\mathrm{i}v_{n})+\sqrt{x_{0}}
\label{eq22},\\
&&f_{m\sigma}(\tau)=\frac{1}{\sqrt{\beta N_{\Lambda}}}{\sum_{\bm{k},n}}e^{\mathrm{i}(\bm{k}\cdot\bm{R}_{m}-\omega_{n}\tau)}f_{\bm{k}\sigma}(\mathrm{i}\omega_{n})
\label{eq23},  
\end{eqnarray}
where $N_{\Lambda}$ denotes the number of unit cells and $x_{0}$ is the condensate density of chargons, the action (\ref{eq19}) can be written in frequency-momentum space as
\begin{eqnarray}
S_{E}&=&{\sum_{\bm{k},n}}'\Psi_{\eta}^{X\dagger}\Big[\Big(\frac{v_{n}^{2}}{2U}+\rho\Big)\delta_{\eta\kappa}+{\cal{H}}_{\eta\kappa}^{X}\Big]\Psi_{\kappa}^{X}\nonumber\\
&&+\sum_{\bm{k},n}\Psi_{\sigma}^{f\dagger}\big[\big(-\mathrm{i}\omega_{n}\big)\delta_{\sigma\sigma^{\prime}}+{\cal{H}}_{\sigma\sigma^{\prime}}^{f}\big]\Psi_{\sigma^{\prime}}^{f}+\cdots.
\label{eq24}
\end{eqnarray}
Some explanations about this formula are given below. $v_{n}=2n\pi/\beta$ is the Matsubara frequencies for bosons field $X$ and $\omega_{n}=(2n+1)\pi/\beta$ for fermions field $f$. The first summation excludes the point $(\mathrm{i}v_{n}^0,\bm{k}^0)$ at which the condensate of chargons occurs. For the $X$-field sector, $\Psi^{X}=\big(X_{\bm{k}}^{A}(\mathrm{i}v_{n}),X_{\bm{k}}^{B}(\mathrm{i}v_{n}),X_{\bm{k}}^{C}(\mathrm{i}v_{n}),X_{\bm{k}}^{D}(\mathrm{i}v_{n})\big)^{T}$ is a $4\times 1$ matrix and Hamiltonian matrices of the $X$-field is
\begin{eqnarray}
{\cal{H}}^{X}=
\begin{pmatrix}
0&\rho_{X}(k_{x})&\rho_{X}(k_{y})&0\\
\rho^{*}_{X}(k_{x})&0&0&\rho_{X}(k_{y})\\
\rho^{*}_{X}(k_{y})&0&0&\rho_{X}(k_{x})\\
0&\rho^{*}_{X}(k_{y})&\rho^{*}_{X}(k_{x})&0
\end{pmatrix},
\label{eq25}
\end{eqnarray}
where $\rho_{X}(k)=Q_{X}(w+ve^{-\mathrm{i}k})+Q_{X}^{\prime}(t_{12}e^{\mathrm{i}k}+t_{21}e^{-\mathrm{i}2k})=|\rho_{X}(k)|e^{\mathrm{i}\phi_{X}(k)}$. For the $f$-field sector, $\Psi^{f}_{\sigma}=\big(f_{\bm{k}\sigma}^{A}(\mathrm{i}\omega_{n}),f_{\bm{k}\sigma}^{B}(\mathrm{i}\omega_{n}),f_{\bm{k}\sigma}^{C}(\mathrm{i}\omega_{n}),f_{\bm{k}\sigma}^{D}(\mathrm{i}\omega_{n})\big)^{T}$ is a $4\times 1$ matrix and the Hamiltonian matrix ${\cal{H}}^{f}$ of the $f$-field is a $2\times 2$ block matrix with ${\cal{H}}^{f}_{\uparrow\downarrow}={\cal{H}}^{f}_{\downarrow\uparrow}=\bm{0}$ and 
\begin{eqnarray}
{\cal{H}}^{f}_{\uparrow\uparrow}\!={\!\cal{H}}^{f}_{\downarrow\downarrow}\!=\!
\begin{pmatrix}
0&\rho_{f}(k_{x})&\rho_{f}(k_{y})&0\\
\rho^{*}_{f}(k_{x})&0&0&\rho_{f}(k_{y})\\
\rho^{*}_{f}(k_{y})&0&0&\rho_{f}(k_{x})\\
0&\rho^{*}_{f}(k_{y})&\rho^{*}_{f}(k_{x})&0
\end{pmatrix},
\label{eq26}
\end{eqnarray}
where $\rho_{f}(k)=Q_{f}(w+ve^{-\mathrm{i}k})+Q_{f}^{\prime}(t_{12}e^{\mathrm{i}k}+t_{21}e^{-\mathrm{i}2k})=|\rho_{f}(k)|e^{\mathrm{i}\phi_{f}(k)}$. 

The Green's functions of chargons and spinons are obtained from the action (\ref{eq24}) as\cite{2015Coleman}
\begin{eqnarray}
G_{X}^{l}(\bm{k},\mathrm{i}v_{n})=\frac{1}{v_{n}^{2}/U+\rho+E_{X}^{l}}.
\label{eq27}\\
G_{f\sigma}^{l_{1}(l_{2})}(\bm{k},\mathrm{i}\omega_{n})=\frac{1}{\mathrm{i}\omega_{n}-E_{f}^{l_{1}(l_{2})}}.
\label{eq28}
\end{eqnarray}
Here, $U$ have been replaced by $U/2$ to preserve the exact atomic limit\cite{2002Florens,2004Florens}. $E_{X}^{l}$ is the lowest energy eigenvalue of Hamiltonian matrices ${\cal{H}}^{X}$ and $E_{f}^{l_{1}(l_{2})}$ are the two lower energy eigenvalues of matrix ${\cal{H}}^{f}_{\uparrow\uparrow}$ or ${\cal{H}}^{f}_{\downarrow\downarrow}$. The poles of Green's function in the imaginary frequency domain determine energy spectra of the bands. Then, the energy spectra of the lower bands of chargons and spinons are $\xi^{l}(\bm{k})=\sqrt{U(\rho+E_{X}^{l})}$  and  $\Xi_{\sigma}^{l_{1}(l_{2})}(\bm{k})=E_{f}^{l_{1}(l_{2})}$ respectively. The derivation of spinon's bands is similar to the one of uncorrelated electrons. So we also label bands of spinons as $(s_1,s_2)$. 

To investigate effects of correlations, the mean field parameters and condensate density $x_{0}$ of chargons or Lagrange multiplier $\rho$ should be determined. We firstly obtain one of the so-called mean field equations from the constraint $|X_{ij}(\tau)|^2=1$. At the mean field level, it is $\sum_{ij}\big<X_{ij}^{*}(\tau)X_{ij}(\tau)\big>=4N_{\Lambda}$. By the Fourier transfroms and noting that the ground state is the eigenstate with the lowest energy, the constraint can be obtained as 
\begin{eqnarray}
\frac{1}{4N_{\Lambda}}{\sum_{\bm{k}}}'\frac{1}{\beta}{\sum_{n}}'G_{X}^{l}(\bm{k},\mathrm{i}v_{n})+x_{0}=1.
\label{eq29}
\end{eqnarray}
Carrying out the Matsubara summation\cite{2015Coleman}, we obtain 
\begin{eqnarray}
\frac{1}{4N_{\Lambda}}{\sum_{\bm{k}}}'\frac{\sqrt{U}}{2\sqrt{\rho+E_{X}^{l}}}+x_{0}=1.
\label{eq30}
\end{eqnarray}
The other mean field equations can be obtained from the definition of mean field parameters (i.e. Eq.~(\ref{eq15})-(\ref{eq18})) as
\begin{eqnarray}
Q_{f}&\!=&\!\frac{-1}{32N_{\Lambda}}{\sum_{\bm{k}}}'\frac{\sqrt{U}\big[g_{X}(k_{x})\!+\!g_{X}(k_{y})\!+\!\text{h.c.}\big]}{2\sqrt{\rho+E_{X}^{l}}}\!+\!x_0
\label{eq31}\\
Q_{f}^{\prime}\!&=&\!\frac{-1}{32N_{\Lambda}}{\sum_{\bm{k}}}'\frac{\sqrt{U}\big[g_{X}^{\prime}(k_{x})\!+\!g_{X}^{\prime}(k_{y})\!+\!\text{h.c.}\big]}{2\sqrt{\rho+E_{X}^{l}}}\!+\!x_0
\label{eq32}\\
Q_{X}\!&=&\!\frac{-1}{8N_{\Lambda}}\left[\sum_{\bm{k}_{2}}g_{f}(k_{y})\!+\sum_{\bm{k}_{3}}g_{f}(k_{x})+\text{h.c.}\right]
\label{eq33}\\
Q_{X}^{\prime}\!&=&\!\frac{-1}{8N_{\Lambda}}\left[\sum_{\bm{k}_{2}}g_{f}^{\prime}(k_{y})\!+\sum_{\bm{k}_{3}}g_{f}^{\prime}(k_{x})+\text{h.c.}\!\right]
\label{eq34}
\end{eqnarray}
Here $g_{X(f)}(k)=(1+e^{-\mathrm{i}k})e^{-\mathrm{i}\phi_{X(f)}(k)}$ and $g_{X(f)}^{\prime}(k)=(e^{\mathrm{i}k}+e^{-\mathrm{i}2k})e^{-\mathrm{i}\phi_{X(f)}(k)}$. In the sum of Eqs.~(\ref{eq33}) and (\ref{eq34}), $\bm{k}_{2}$ belongs to the Brillouin zone where band $(s_1=1,s_2=-1)$ is lower than band $(s_1=-1,s_2=1)$, while $\bm{k}_{3}$ belongs to the Brillouin zone where band $(s_1=-1,s_2=1)$ is lower than band $(s_1=1,s_2=-1$).

\subsection{\label{sec3-2}Influences of the correlation on topological phase diagrams}

\begin{figure}[ht]  
\includegraphics[width=6.5cm,height=4.5cm]{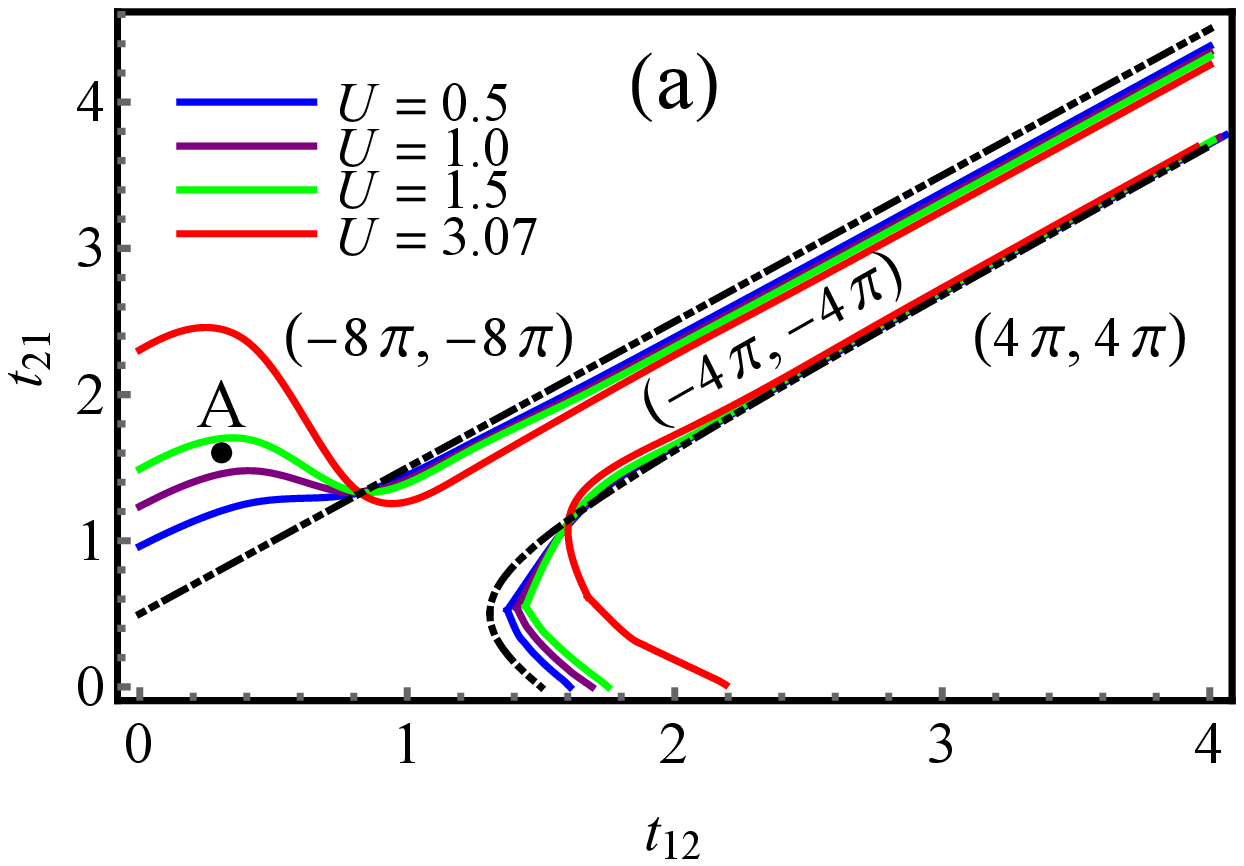}
\includegraphics[width=6.5cm,height=4.5cm]{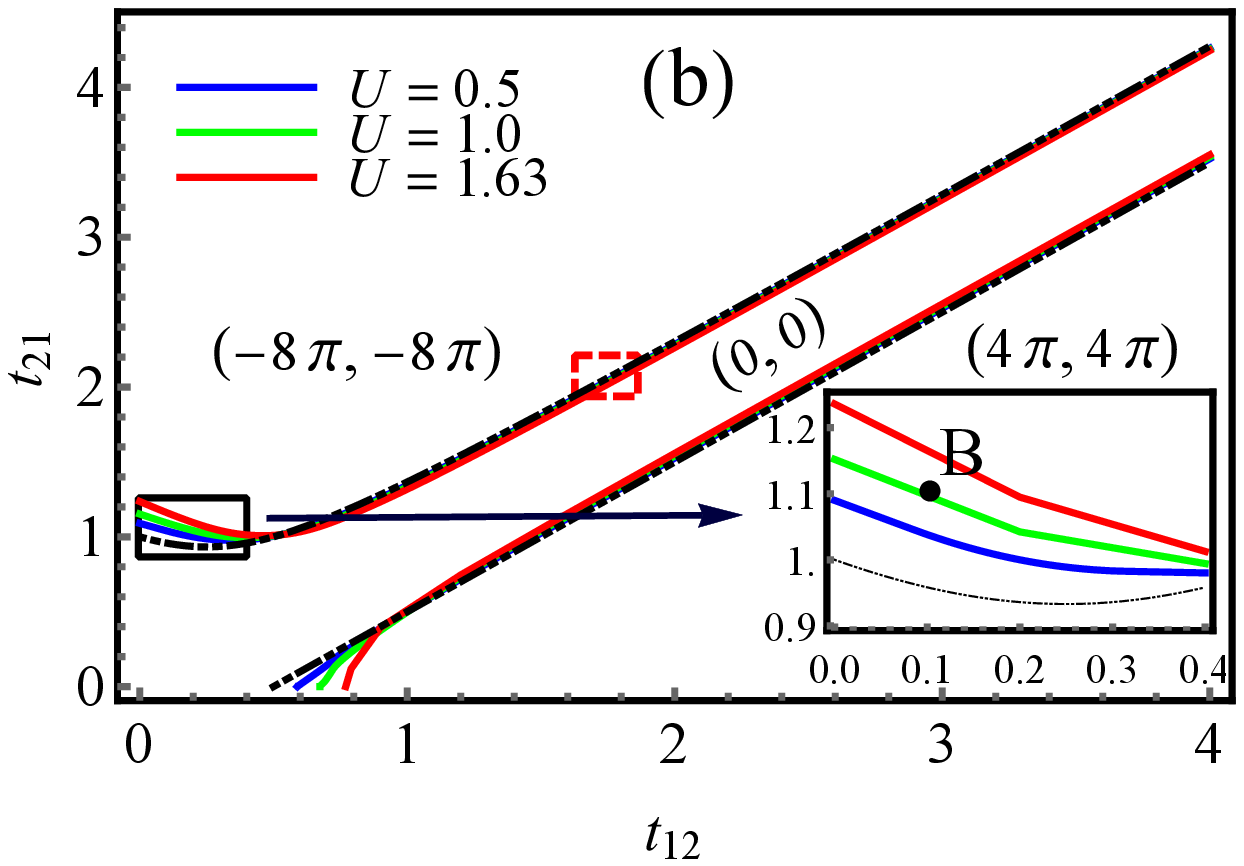}
\includegraphics[width=6.5cm,height=4.5cm]{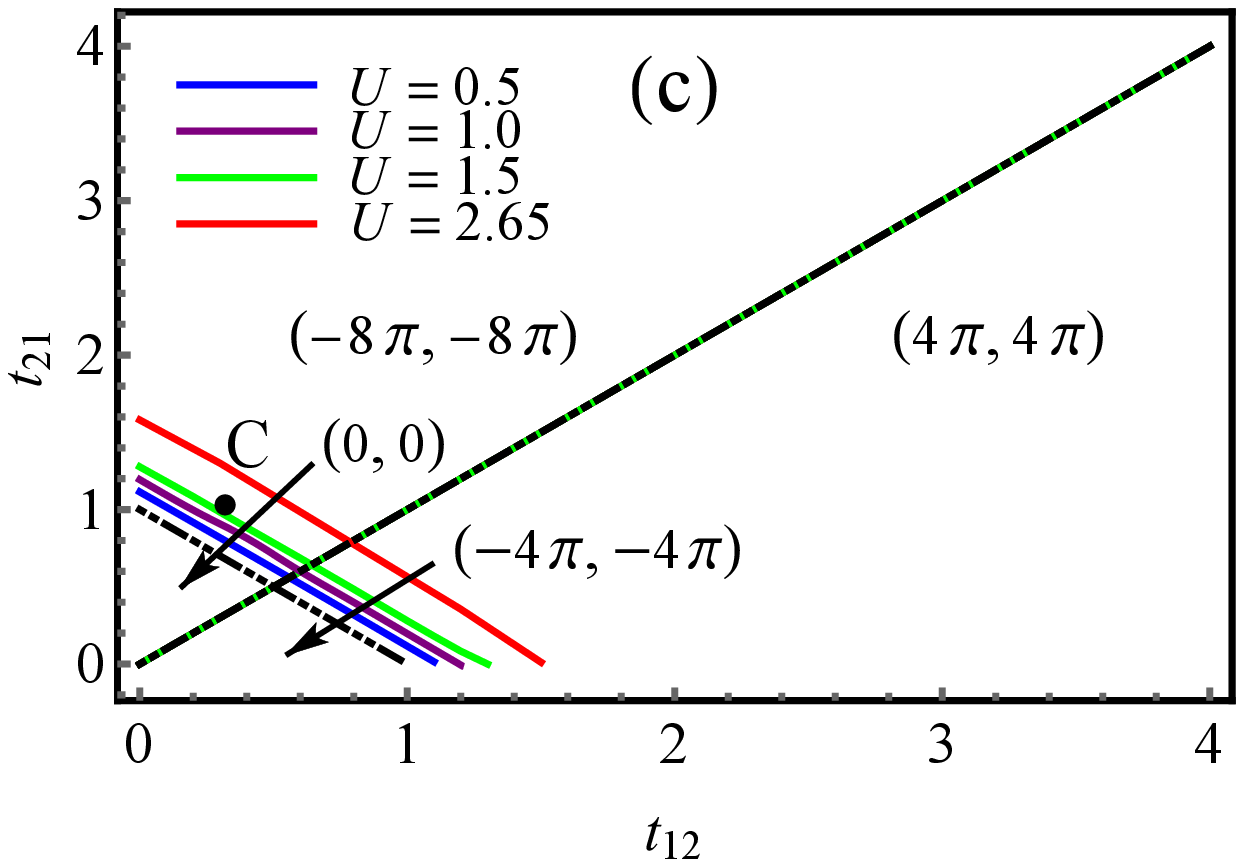}
\caption{\label{fig4}(Color online)Boundaries of topological transitions and phase digrams under correlations. (a) $w=1, v=1.5$, (b) $w=1, v=0.5$ and (c) $w=1, v=1$. Black dashed lines are the boundaries of topological transitions in the case of the absence of correlations. Red lines are the phase boundaries corresponding to the minimum critical $U$ at which the Mott transition of chargons occurs.}
\end{figure}

\begin{figure}[ht]  
\includegraphics[width=6.5cm,height=4.5cm]{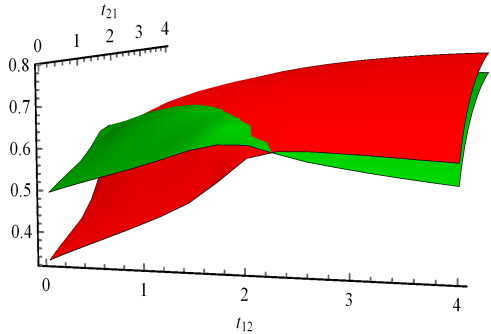}
\caption{\label{fig5}(Color online)Behaviors of $Q_f$ and $Q_{f}^{\prime}$ in the case of $w<v$ ($w=1, v=1.5)$ and $U=1.5$. The green (red) surface represents $Q_{f}$ ($Q_{f}^{\prime}$). Behaviors of $Q_f$ and $Q_{f}^{\prime}$ in the case of $w>v$ and $w<v$ are very similar to the case of $w<v$, so we do not show them here.}
\end{figure}

For the larger correlations $U$ there is a Mott transition of chargons, above which the spin-charge separation occurs and the chargon is uncondensed. In our investigations, $U$ is restricted to the region of the relatively small correlations to guarantee that the chargon is condensed, since we focus on effects of correlations on the topological transitions of physical electrons. To the physical electron or condensed chargons, $x_0\not=0$ and $\rho=-\text{min}(E_{X}^{l})$ which is derived from  $\xi^{l}(\bm{k})=0$. The boundaries of topological transitions of physical electrons can be obtained from the Hamiltonian (\ref{eq26}) of spinons which is similar to the one of uncorrelated electrons but with renormalized electron hoppings $Q_{f}w$, $Q_{f}v$, $Q_{f}^{\prime}t_{12}$ and $Q_{f}^{\prime}t_{21}$. That is 
\begin{eqnarray}
Q_{f}(w-v)+Q_{f}^{\prime}(-t_{12}+t_{21})=0,
\label{eq35}\\
Q_{f}(w+v)+Q_{f}^{\prime}(t_{12}+t_{21})=0,
\label{eq36}
\end{eqnarray} 
and
\begin{eqnarray}
t_{21}(Q_{f}w-Q_{f}^{\prime}t_{21})=t_{12}(Q_{f}v-Q_{f}^{\prime}t_{12}).
\label{eq37}
\end{eqnarray}
These equations can be calculated self-consistently with the mean field equations (\ref{eq30})-(\ref{eq34}). The boundaries of topological transition and phase digrams are showed in Fig. (\ref{fig4}).
In the region of the small $t_{12}$ and $ t_{21}$, the correlation has the larger influence on the boundary of topological transition. However, the effects of correlations on the topological transition become negligible when $t_{12}$ and $t_{21}$ are sufficiently large. It originates from the weak effect of renormalization of correlation on the model parameters, because the mean field parameter $Q_{f}^{\prime}/Q_{f}$ is almost equal to unity in the case of large $t_{12}$ and $t_{21}$. The tendency of $Q_{f}^{\prime}/Q_f\sim 1$ makes the boundries (\ref{eq35}) and (\ref{eq37}) to be almost equivalent to boundries (\ref{eq8}) and (\ref{eq10}) in the case of uncorrelated model, respectively. The behaviors of $Q_f$ and $Q_{f}^{\prime}$ with the increasing $t_{12}$ and $t_{21}$ are shown in Fig.~\ref{fig5}. 
Especially, to the case of $w=v$, Eq.~(\ref{eq35}) is exactly Eq.~(\ref{eq8}), i.e. $t_{12}=t_{21}$. The effect of renormalization of correlations on the boundary (\ref{eq35}) of topological transitions can't be acquired in this mean field method. The situation is show in Fig.~\ref{fig4}(c). The vectored Berry phase of each part of the diagrams is also obtained from the winding number of $\rho_{f}(k_l)$ when it winds around the origin of the complex plane. It is obvious that the winding pattern of $\rho_{f}(k_{l})$ is similar to the one of $\rho(k_{l})$, because the $w$, $v$, $t_{12}$ and  $t_{21}$ in $\rho(k_{l})$ are just renormalized by correlations to new values multiplied by $Q_{f}$ or $Q_{f}^{\prime}$ whose behaviors are illustrated in Fig.~\ref{fig5}.

\begin{figure}[ht]  
\includegraphics[width=6.5cm,height=4.5cm]{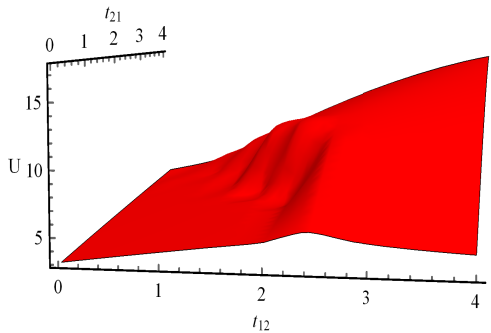}
\caption{\label{fig6}(Color online)The boundary of the Mott transition for chargons in the case of $w<v$ ($w=1, v=1.5$). The minimum $U_{c}\approx 3.07$ for Mott transition occurs at $t_{12}=t_{21}=0$. For cases of $w>v$ and $w=v$, the critical U have the same trend in their growth as the case of $w<v$. So we just draw the boundary when $w<v$. }
\end{figure}

In the slave-rotor mean field method, there is a spin-charge seperation beyond the Mott transition of  chargons. The boundary of the Mott transition in the case of $w<v$ is shown in Fig.~\ref{fig6}. 
We enforce the Hubbard strength satisfy $U<U_{c}$, where $U_{c}\approx 3.07$ is the minimum critical value for the Mott transition in the case of $w=1$ and $v=1.5$, because we focus on the phase of physical electrons. The minimum critical value of U also occur at $t_{12}=t_{21}=0$ for $w>v$ and $ w=v$, and are approximately equal to 1.63 and 2.65 for $w=1,v=0.5$ and $w=v=1$ respectively. The phase boundary corresponding to $U_c$ has been shown in Fig.~\ref{fig4} for the three cases.  When $U>U_c$, we can also obtain boundaries of topological transitions of physical electrons, but phases of spinons (not physical electrons) will emerge in some regions of phase diagrams. We leave the investigations on this case in this work.      

\subsection{\label{sec3-3}Correlations-driven topological transitions}

The shift of phase boundaries caused by correlations indicates the correlation-driven topological transition. For example, there are topological transitions of phase $(-8\pi,-8\pi)\rightarrow$ phase $(-4\pi,-4\pi)$, phase $(4\pi,4\pi)\rightarrow$ phase $(-4\pi,-4\pi)$, phase $(-4\pi,-4\pi)\rightarrow$ phase $(-8\pi,-8\pi)$ and phase $(-4\pi,-4\pi)\rightarrow$ phase $(4\pi,4\pi)$ in the case of $w<v$. To gain some insights into the correlations-driven topological transition of this model, let us consider a few special transitions below.

\begin{figure}[ht]  
\includegraphics[width=6.5cm,height=4.5cm]{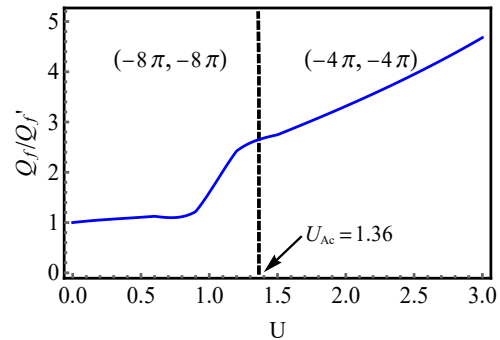}
\caption{\label{fig7}(Color online)The correlations-driven topological transition for the point A in Fig.~\ref{fig4}(a) in the case of $w=1, v=1.5$. When the increasing U brings the ratio $Q_{f}/Q_{f}^{\prime}=2.6$, the topological transition occurs.}
\end{figure}

\begin{figure}[ht]  
\includegraphics[width=4.25cm,height=4cm]{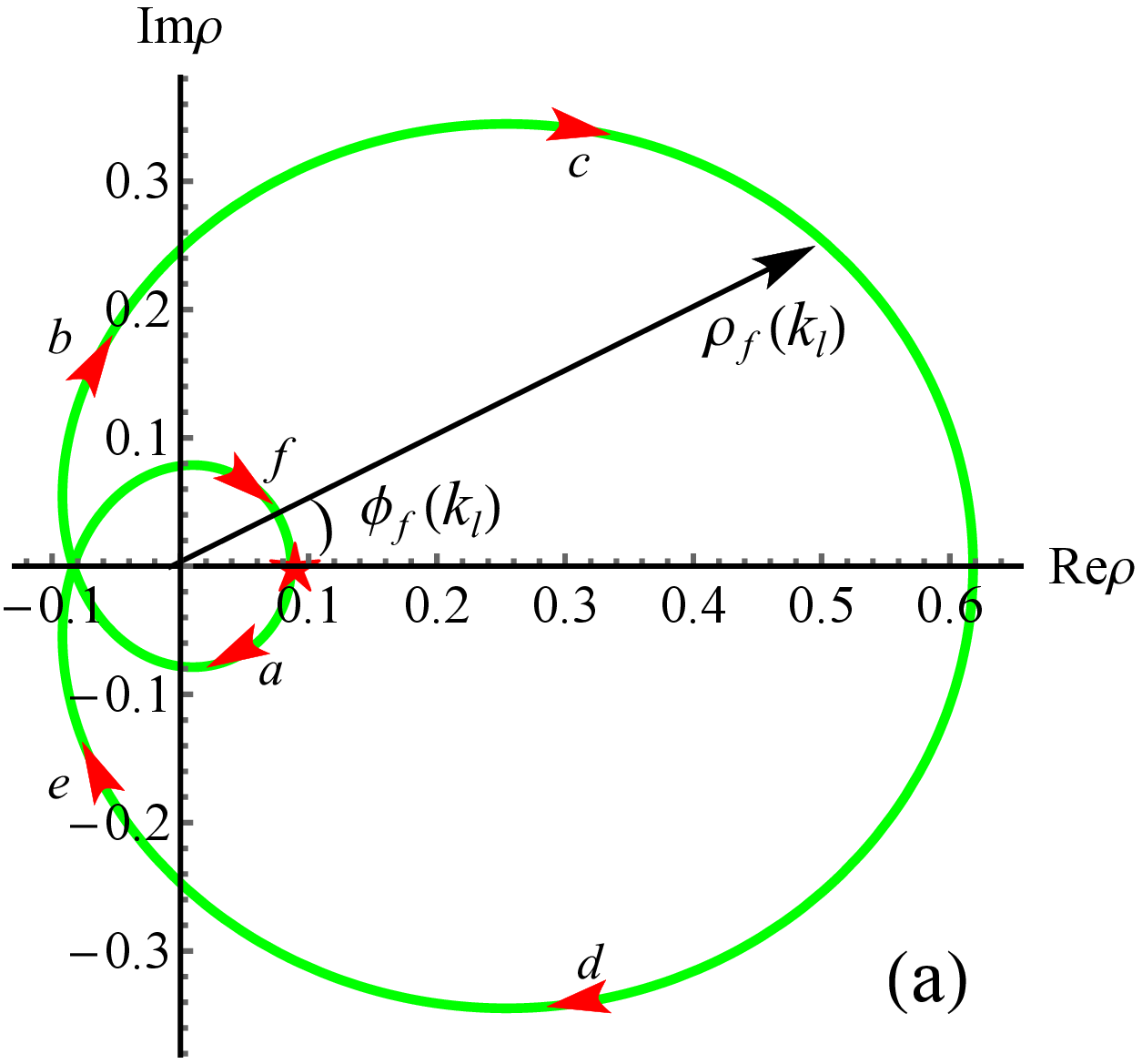}
\includegraphics[width=4.25cm,height=4cm]{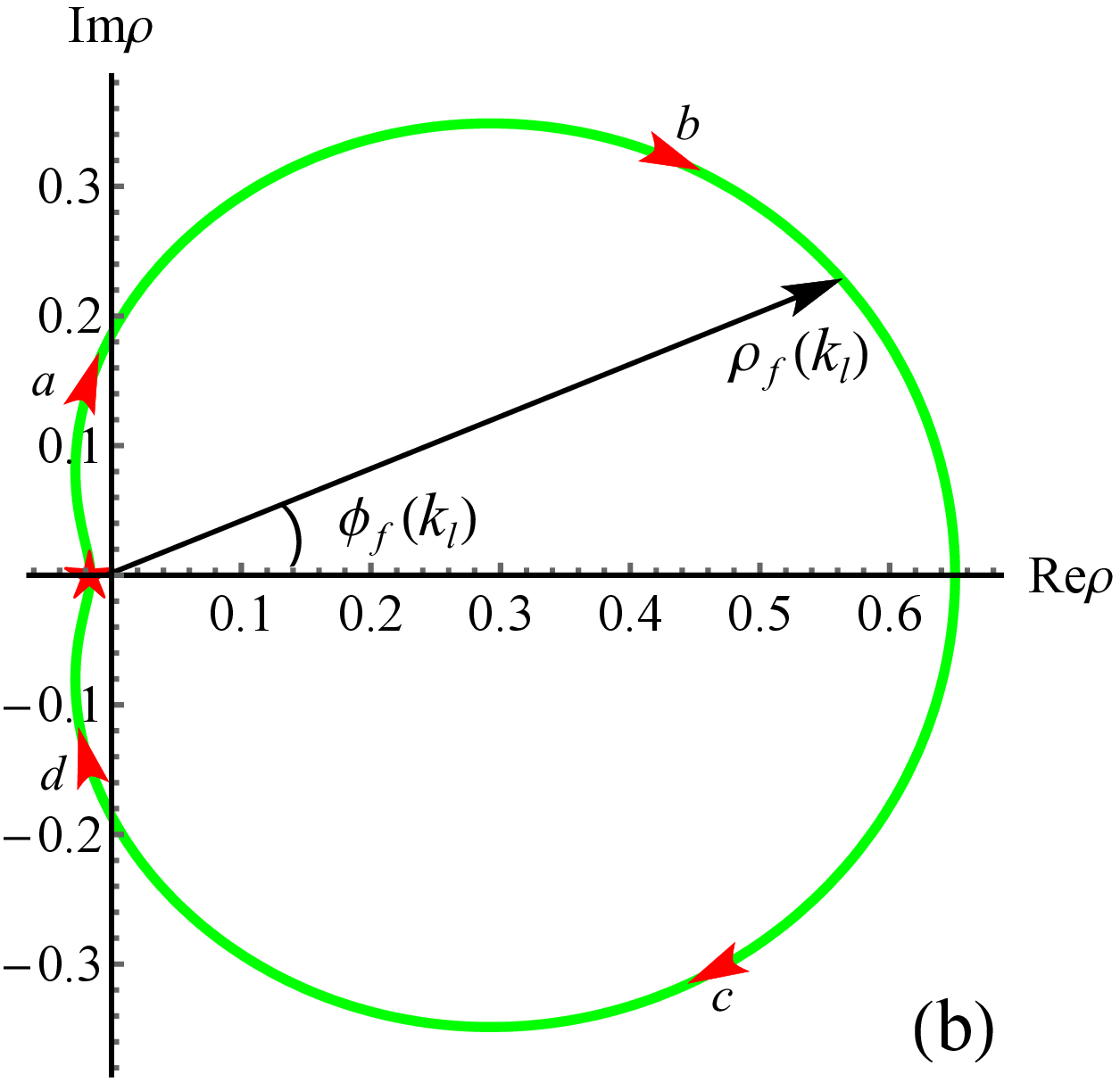}
\caption{\label{fig8}(Color online)The winding pattern of $\rho_{f}(k_l)$ in the case of $t_{12}=0.3,  t_{21}=1.6$ and $w=1, v=1.5$. (a) $U=0.8 (<U_{Ac})$; (b) $U=1.8 (>U_{Ac})$.}
\end{figure}

For the case of $w<v$, we choose the point $(t_{12},t_{21})=(0.3,1.6)$ in the phase diagram (point A in Fig.~\ref{fig4}(a)). When $U=0$, the system stays in the topological phase $(-8\pi, -8\pi)$. From Eq.~(\ref{eq35}), if the increasing $U$ brings the ratio $Q_{f}/Q_{f}^{\prime}$ to the critical value $(t_{12}-t_{21})/(w-v)$, the topological transition from phase $(-8\pi, -8\pi)$ to phase $(-4\pi, -4\pi)$ occurs. The topological transition are illustrated in the Fig.~\ref{fig7}.
The critical value of Hubbard strength $U_{Ac}$ is roughly $1.36$ in the case of $w=1, v=1.5$ and the two topologically non-trivial phases are verified by the winding pattern of $\rho_{f}(k_{l})$ as shown in Fig.~\ref{fig8}.    

\begin{figure}[ht]  
\includegraphics[width=6.5cm,height=4.5cm]{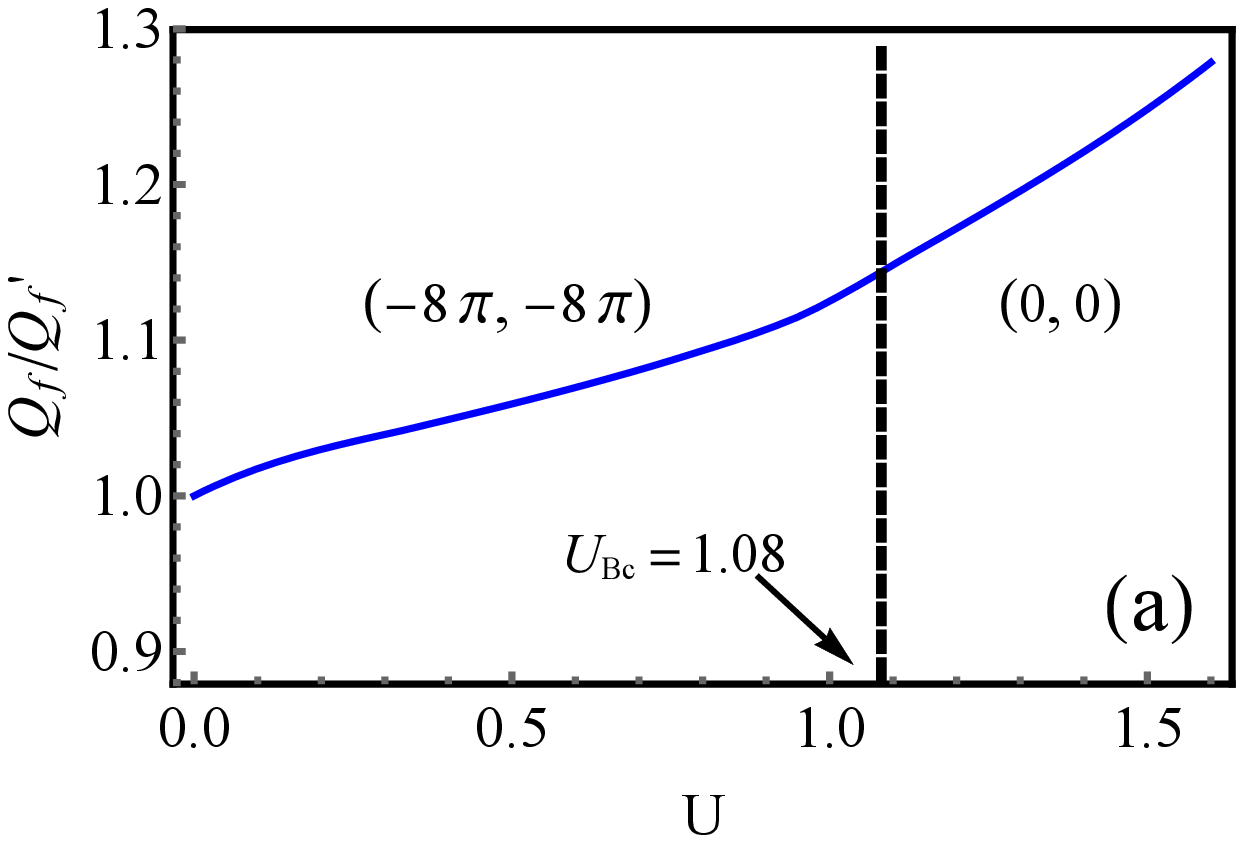}
\includegraphics[width=6.5cm,height=4.5cm]{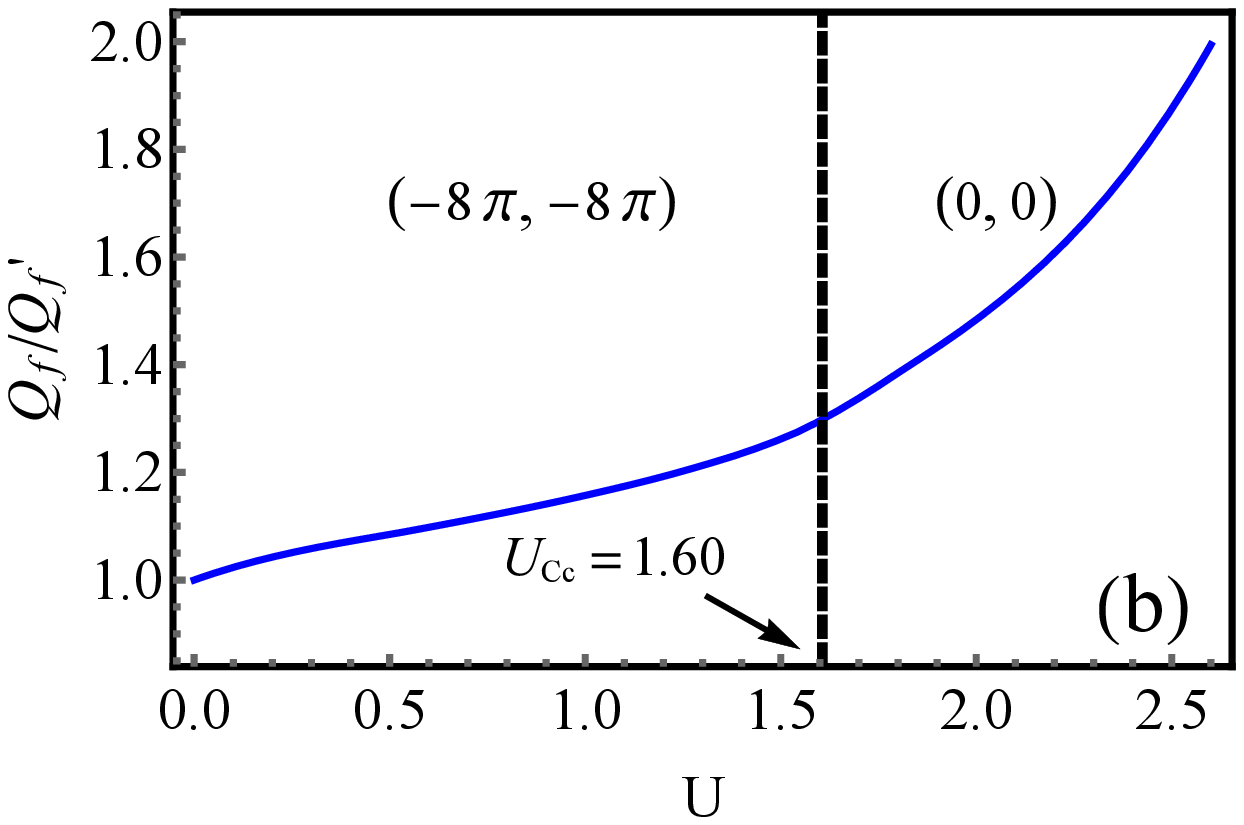}
\caption{\label{fig9}(Color online)The topological transitions driven by correlations. (a) the case of $w=1, v=0.5$ and $t_{12}=0.1, t_{21}=1.1$. (b) the case of $w=1, v=1$ and $t_{12}=0.3, t_{21}=1$.}
\end{figure}

For $w>v$, the point $(t_{12},t_{21})=(0.1,1.1)$ (point B in Fig.~\ref{fig4}(b)) in the phase diagram is chosen to illustrate the topological transition in this case. Now the critical Hubbard strength should enfore the ratio $Q_{f}/Q_{f}^{\prime}$ to satisfy Eq.(\ref{eq37}), i.e. $Q_{f}/Q_{f}^{\prime}=(t_{21}^{2}-t_{12}^{2})/(wt_{21}-vt_{12})$. The topological transition from phase $(-8\pi, -8\pi)$ to phase $(0, 0)$ is shown in Fig.~\ref{fig9}(a). The critical $U_{Bc}$ is roughly 1.08 and $Q_{f}/Q_{f}^{\prime}\approx 1.14$ is satisfied. The point $(t_{12},t_{21})=(0.3,1)$ (point C in Fig.~\ref{fig4}(c)) is chosen for $w=v=1$. The topological transition which is constrained by Eq.~(\ref{eq37}), i.e. $Q_{f}/Q_{f}^{\prime}=1.3$, and occurs at $U_{Cc}\approx 1.60$ is shown in Fig.~\ref{fig9}(b). The two correlations-driven topological transitions can also be confirmed by the winding pattern of $\rho_{f}(k_l)$ as the case of $w<v$.

\begin{figure}[ht]  
\includegraphics[width=6.8cm,height=4.5cm]{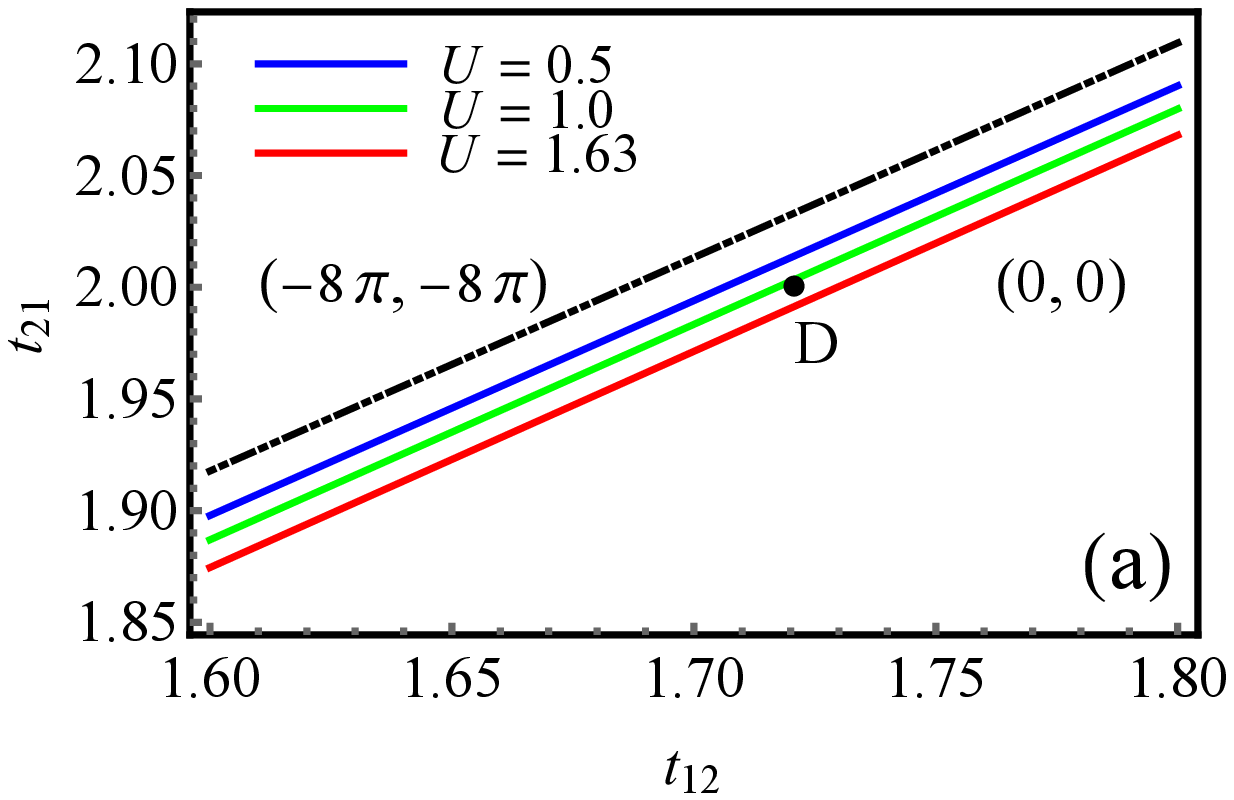}
\includegraphics[width=6.5cm,height=4.5cm]{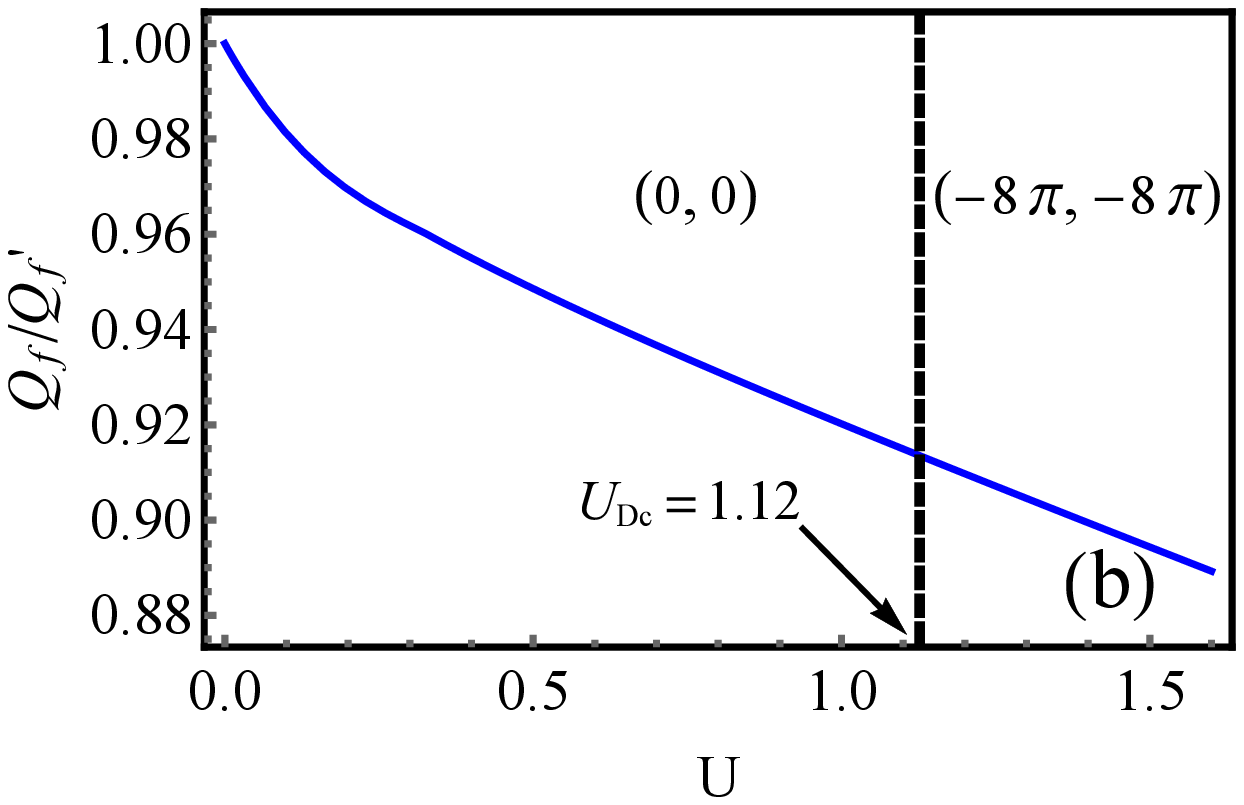}
\caption{\label{fig10}(Color online)(a) A part of the narrow regions of phase diagram (Fig.~\ref{fig4}(b)), which is marked by the red dashed rectangle in Fig.~\ref{fig4}(b). (b) The correlations-driven topological transition from phase $(0, 0)$ to phase $(-8\pi, -8\pi)$ in the case of $w=1, v=0.5$ and $t_{12}=1.72, t_{21}=2$ (point D in (a)).}
\end{figure}

The more interesting phenomenon may be the emergence of the topological phase driven by correlations from a topologically trivial phase. In the case of $w>v$, there are two narrow windows in the phase diagram (Fig.~\ref{fig4}(b)) for topological transition from topologically trivial phase $(0,0)$ to topologically non-trivial phase $(-8\pi,-8\pi)$ and $(4\pi, 4\pi)$ respectively. Let us focus on the former. We amplify a part of the narrow region in the phase diagram of $w=1, v=0.5$ as shown in Fig.~\ref{fig10}(a). When $U=0$, the system at point $(t_{12},t_{21})=(1.72, 2)$ (point D) posseses topologically trivial phase $(0, 0)$. If the increasing Hubbard U reaches the critical value $U_{Dc}\approx 1.12$, Eq.~(\ref{eq37}), i.e. $Q_{f}/Q_{f}^{\prime}\approx 0.91$, is satisfied and the topological transition to phase $(-8\pi, -8\pi)$ occurs, as shown in Fig.~\ref{fig10}(b).

\section{\label{sec4}conclusions and outlook}
Although the Chern number and the Berry curvature vanish, the two-dimensional SSH model also possesses topologically non-trivial phases which are characterized by the so-called vectored Berry phase. Like the case of one-dimensional SSH model, the additional third nearest neighbor hoppings can also drive the topological transitions and enlarge the topological invariant. It is actually a general phenomenon that also emerges in the two-dimensional Chern insulators where the longer-range hoppings lead to the larger Chern number. The phase boundaries are obtained from the critical situations that the $\rho(k_l)$ passes through the origin of the complex plane and the vectored Berry phase is obtained from the winding pattern of $\rho(k_l)$. We give phase diagrams in all of the cases of nearest neighbor hoppings, i.e. $w<v$, $w>v$ and $w=v$.

Correlations can shift these phase boundaries of non-interacting phase diagrams, especially in the domains of the maller third nearest neighbor hoppings, and then lead to topological transitions. Our investigations from the winding number show that there are many correlations-driven topological transitions from phases with the larger topological invariant to that with the smaller topological invariant, and vice versa (see Fig.~\ref{fig4}).  More interestingly, there are two correlations-driven topologically non-trivial phase in the case of $w>v$, i.e. phases $(-8\pi, -8\pi)$ and $(4\pi, 4\pi)$ from the trivial phase $(0, 0)$ , although the region for the two transitions in the interacting phase diagram is very narrow. Finally, what we want to do is investigate the effect of correlations on physical electron, so the strength of correlations is restricted to avoid the Mott transition of charge degree of freedom.

The intrinsic and extrinsic (e.g. Rashba) SOCs have profound influences on the bands structure of topological insulators \cite{2014Laubach,2020Du}. For SSH model, it can partly be seen in the case of one dimension and intrinsic SOC \cite{2014Yan,2016Bahari}. It is worthwhile to consider effects of the two SOCs (furthermore, the interplay of correlations and SOCs) in the two-dimensional SSH model with/without the longer-range hoppings. For the extensions of SSH model to other two-dimensional lattices (e.g. honeycomb lattice), the mechanism to acquire a larger topological invariant by the longer-range hoppings should work. But it has not yet been confirmed. Furthermore, the effect of correlations on these two-dimensional SSH models may be an interesting issue.

\begin{acknowledgments}
This work was supported by National Natural Science Foundation of China under Grant No. 11964042.
\end{acknowledgments}

\end{document}